\documentclass[fleqn,usenatbib]{mnras}

\usepackage{newtxtext,newtxmath}

\usepackage[T1]{fontenc}
\usepackage{ae,aecompl}

\usepackage{graphicx}	
\usepackage{amsmath}	
\usepackage{float}

\newcommand{\noopsort}[2]{#2}

\title[Carbon white dwarfs]{The non-explosive stellar merging origin of the ultra-massive carbon-rich white dwarfs}

\author[A. Kawka et al.]{
Adela Kawka$^{1}$\thanks{Contact e-mail: \href{mailto:adela.kawka@curtin.edu.au}{adela.kawka@curtin.edu.au}}, 
Lilia Ferrario$^{1, 2}$ and St\'ephane Vennes$^{1, 2}$
\\
$^1$ International Centre for Radio Astronomy Research - Curtin University, GPO Box U1987, Perth, WA 6845, Australia\\
$^2$ Mathematical Sciences Institute, The Australian National University, Canberra, ACT 0200, Australia}

\date{Accepted XXX. Received YYY; in original form ZZZ}

\pubyear{2022}

\begin{document}
\label{firstpage}
\pagerange{\pageref{firstpage}--\pageref{lastpage}}
\maketitle

\begin{abstract}
We have investigated the origin of a sub-class of carbon-polluted white dwarfs (DQ) 
originally identified as the ``hot DQ" white dwarfs.
These objects are relatively hot ($10\,000\lesssim T_{\rm eff}\lesssim25\,000$\,K), have markedly higher carbon abundance (C-enriched), are more massive ($M\gtrsim0.8$\,M$_\odot$) than ordinary DQs ($M\sim0.6$\,M$_\odot$), and display high space velocities. Hence, despite their young appearance their kinematic properties are those of an old white dwarf population. The way out of this dilemma is to assume that they formed via the merging of two white dwarfs. In this paper we examine the observed characteristics of this population of ``C-enriched" DQ white dwarfs and confirm that nearly half of the 63 known objects have kinematic properties consistent with those of the Galactic thick disc or halo. We have also conducted population synthesis studies and found that the merging hypothesis is indeed compatible with observations. Studies of this sub-class of white dwarfs have important implications for our understanding of Type Ia Supernovae (SNeIa), commonly used to determine the expansion history of the universe, since the same formation channel applies to both kind of objects. Hence probing the properties of these white dwarfs that failed to explode may yield important constraints to the modelling of the mechanisms leading to a thermonuclear runaway.

\end{abstract}


\section{Introduction}

White dwarfs are the final stage of stellar evolution for the majority of stars and about a 
quarter of white dwarfs are found in binary systems \citep{Hollands2018}. However, about 
half of intermediate mass main-sequence stars (the progenitors of the white dwarfs) are 
in binaries \citep{Ferrario2012, Duchene2013}. This discrepancy provides evidence that some white dwarfs formed in stellar mergers \citep{Briggs2015, Toonen2017}.

The merger of double degenerates (DDs) provides one path toward Type Ia supernova explosions. 
The study of white dwarfs that lead to such events, but failed to explode, yield important 
constraints to the modelling of the mechanisms leading to a thermonuclear runaway \citep{Ruiter2020}.
The expectation is that white dwarfs formed via DD mergers are more 
massive than white dwarfs descending from single stars and this was indeed shown to be the case by the population synthesis calculations of \citet{Briggs2015, Briggs2018a}.
\citet{Temmink2020} further investigated the impact of binary evolution on apparently single
white dwarfs and found that about a quarter of them should have formed in
a stellar merger, but that only about 3 per cent formed from a DD merger, which is consistent with the findings of \citet{Briggs2015}.
\citet{sch2021} calculated evolutionary models for these white dwarf merger remnants and showed that they should be massive and fast rotating.
Overall, the products of DD mergers should be rare.

The entire sub-class of hot, carbon-rich white dwarfs are candidate merger products \citep{Dunlap2015}.
White dwarfs with carbon lines or molecular carbon bands are generally classified as DQs \citep{Dufour2005}. The presence of carbon in most of 
these stars can be explained by the dredge-up of core carbon by a deep helium convection zone \citep{Pelletier1986,fon2005}.
This model explains the observed increase in the carbon abundance with increasing temperatures up to $T_{\rm eff} \lesssim 10\,000$\,K but fails to explain the surge in carbon observed in hotter DQ white dwarfs \citep{Dufour2007, Dufour2008, Coutu2019}.
These hot DQ white dwarfs display neutral and ionized carbon line spectra in a carbon dominated atmosphere at effective temperatures ranging from $\sim 18\,000$ to $25\,000$~K \citep{Dufour2008}. Cooler objects display a blend of neutral carbon lines and molecular carbon bands at temperature ranging from $\sim 10\,000$ to $16\,000$\,K \citep{Coutu2019} in a mixed carbon/helium atmosphere.
Using Gaia parallaxes, \citet{Coutu2019} suggested that these hot DQs should also be very massive ($\gtrsim 0.8$\,M$_\odot$). Furthermore, \citet{Dunlap2015} showed that these hot DQs have high transverse velocities. 
This means that despite their high temperatures and high masses, characteristic of a young white dwarf population, they would in fact belong to an old one. These properties, along with the presence of a magnetic field and high rotation rate \citep{Dunlap2015}, strongly suggest that these objects formed through the merger of two white dwarfs.

This population of massive carbon-polluted merger products should extend below $\sim10\,000$\,K and could be distinguished
from canonical DQ white dwarfs by having a higher carbon abundance at a given temperature.  
Therefore, we are dealing with an unusual white dwarf population that has a different origin from the general white dwarf population.
To avoid confusion, we shall refer to the combined population of hot DQ white dwarfs and their cooler counterparts as ``C-enriched DQs''.

Strong independent evidence in support of the merger hypothesis for the origin of the C-enriched DQs is 
provided by a kinematical study of the DQ white dwarf LP~93-21. \citet{Kawka2020LP93} showed that the total age of LP~93-21, 
assuming single star evolution, is too short and inconsistent with its Galactic kinematics which place LP~93-21 in the halo hence bolstering a merger origin for the star.

In this paper we extend the kinematical approach to the now much larger sample of massive DQ white dwarfs and we investigate, through a population synthesis study, whether the stellar merger hypothesis for the origin of the C-enriched DQs is consistent with observed population properties (age, mass).

\section{Massive DQ white dwarfs}

We have gathered all known DQ white dwarfs from several sources including those from the recent studies by \citet{Coutu2019}, \citet{Blouin2019} 
and \citet{Koester2019} and we have adopted the properties (effective temperature, surface gravity, mass and carbon abundance) from these
sources.
For stars that have not been analysed using Gaia measurements, we used available spectroscopy, photometric measurements and Gaia 
distance measurements to determine their properties. If the effective temperature and carbon abundance are known, we combined this information with the Gaia parallax and photometry and our model photometry to estimate the surface gravity and mass. We used the 
evolutionary mass-radius relations for helium atmospheres of \citet{Benvenuto1999} combined with our model DQ spectra 
\citep{Kawka2020LP93} to calculate absolute magnitudes.

\begin{figure}
\includegraphics[viewport=30 325 565 690,clip,width=0.99\columnwidth]{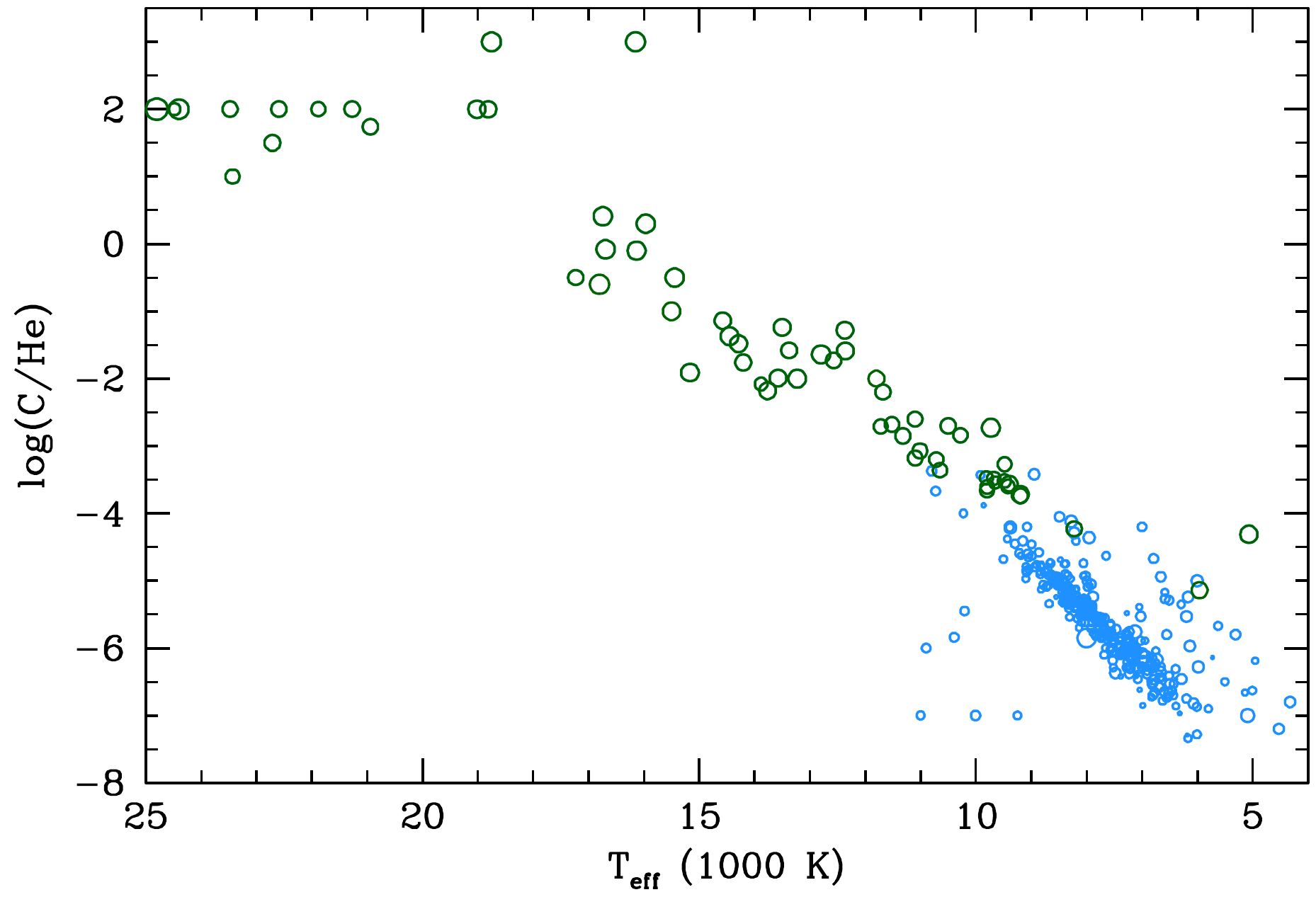}
\caption{Carbon abundance measurements as a function of the effective temperature of the known population
of DQ white dwarfs. The massive C-enriched DQs are shown in green whilst the others are shown in blue. The size of the circles is proportional to the mass of the white dwarf.
\label{fig_abun}}
\end{figure}

From this sample we extracted white dwarfs with masses greater than 0.8\,M$_\odot$ that we consider representative of potential merger products. We checked all massive candidates for their carbon abundance to ensure they lie in the enhanced carbon sequence and removed those that appear to lie in the lower carbon abundance sequence. 
To this sample we added two new massive DQs from the SkyMapper survey of high proper motion white dwarfs (Vennes et al., in preparation, see Appendix A).
The spectra of these two stars were obtained with the Focal Reducer and low dispersion Spectrograph (FORS) at the European Southern
Observatory and the Wide-Field Spectrograph (WiFeS) at Siding Spring Observatory (SSO).
These massive and C-enriched DQs are listed in Table\,\ref{tab_prop_massive} with their atmospheric parameters and they are shown as the green circles in Fig.~\ref{fig_abun}.

\subsection{Kinematics}\label{kinematics2}

We calculated the Galactic velocity components using the distance and proper motion from Gaia, and, wherever possible, a measurement of 
the radial velocity. For most stars we used spectra from the Sloan Digital Sky Survey (SDSS). In a few cases we extracted archival spectra from the European Southern Observatory (ESO) Archive and the Keck Observatory Archive: A UVES spectrum of J0045$-$2336 obtained on 2016 Sep 4 (Programme ID 097.D-0063), a series of PMOS (FORS2) spectra of J0106$+$1513, J0236$-$0734, and J0818$+$0102 obtained on 2012 Nov 17-18 (Programme 090.D-0536), and Low Resolution Imaging Spectrometer (LRIS) spectrum of J0205$+$2057 obtained on 1996 Dec 12. We also obtained a FORS2 spectrum of J2255$-$2826 on 2016 Jun as part of our ESO Programme 097.D-0694, and spectra of J1225$+$0959 and J2140$-$3637 obtained at Siding Spring Observatory (SSO) with the 2.3-m telescope and the Wide-Field Spectrograph (WiFeS) on 2020 Mar 19 and 2015 Sep 25, respectively.

Interestingly, the high resolution UVES spectra of J0045-2336 revealed that some lines show strong red asymmetry, similar to \ion{Si}{ii} 
lines observed in the heavily polluted GALEX~J193156.8+011745 \citep{Vennes2011}. At low resolution, these asymmetric profiles appear as redshifted profiles as noted in the analysis presented by \citet{Kawka2020LP93}.
In the high electronic density of massive DQs, the carbon line positions are altered by the Stark effect which induces large radial velocity
shifts \citep[see a discussion in][]{Kawka2020LP93}. 
In cooler objects dominated by neutral carbon, the extent of this shift ($d_e$) depends on the electronic density in the 
atmosphere and, given the range of temperatures and masses among these objects, the Stark shift was estimated at $d_e\approx-15$\,km\,s$^{-1}$ for measurements based on the
spectral line \ion{C}{i}$\,\lambda5181.83$\AA\ which is only minimally Stark shifted. In hotter objects, dominated by singly ionized carbon, the total Stark shift is estimated at $d_e+d_{\ion{C}{ii}}\approx45$\,km\,s$^{-1}$ for measurements based on \ion{C}{ii}$\,\lambda 4270$\AA\ \citep{lar2012}. These velocity measurements were corrected 
for the Stark shift as well
as for the gravitational redshift. All measurements are barycentric corrected. In the presence of a magnetic field, the central ($\pi$) component was employed for the measurement.
For seventeen stars it was not possible to measure a radial velocity because of
a low signal-to-noise ratio or a lack of publicly available spectra. For these stars we assumed a radial velocity of zero.

The Galactic 
space velocity components were computed using \citet{JohnsonSoderblom1987} and assuming that the Solar motion relative to the local 
standard of rest is ($U_\odot,V_\odot,W_\odot$) = (11.1, 12.2, 7.3) km~s$^{-1}$ \citep{Schonrich2010}. The resulting velocity vectors 
for the C-enriched DQs are listed in Table\,\ref{tbl_kin}.
We have also derived the $z-$component of their angular momenta, $J_z$, and the eccentricity
of their orbit, $e$ using \textsc{galpy} \citep{bov2015}. We assumed that 
the rotational speed of the Galactic disc is 220\,km\,s$^{-1}$ and the 
Sun's distance from the Galactic centre is 8~kpc. These kinematical properties provide 
an additional constraint to help distinguish between thin disc, thick disc and halo stars \citep{Pauli2003,Pauli2006}.

\subsection{Space density of C-enriched DQ white dwarfs}\label{SpaceDensity}

\citet{Hollands2018} estimated a white dwarf space density of 
$4.49\pm0.38 \times 10^{-3}$~pc$^{-3}$ using the 20~pc sample. In this sample 75 per cent are isolated white dwarfs, that is 104
white dwarfs. Using population syntheses \citet{Temmink2020} showed that 10 to 30 per cent of observed single white dwarfs would have formed through binary
mergers, and of these 15 per cent would have formed via DD mergers. Therefore within 20~pc, about 10 to 31 single
white dwarfs would have formed in binary mergers and of these, 1 to 5 would have formed in DD mergers for a space density of 7-$35\times10^{-5}$~pc$^{-3}$ . The C-enriched white dwarfs would only be a subset of all merger products alongside other merger candidates such as ultra-massive magnetic white dwarfs \citep[e.g., RE/EUVE~J0317$-$855, ][]{Ferrario1997,Vennes2003}. 
We can estimate the space density of C-enriched DQs by bracketing its value between a maximum value based on a complete, local sample, and a minimum value based on a list of all known objects. There are no C-enriched DQs within 20~pc allowing us to place an upper limit to their space density at $\la3\times10^{-5}$~pc$^{-3}$. Two C-enriched DQs (G47-18 and SMSSJ2140-3637) are found within 40~pc. In the same volume \citet{mccleery2020} reported 1233 white dwarf candidates from the Gaia DR2 \citep{Gen2019}, out of which a minimum of 521 ($\approx 40$ percent) were spectroscopically confirmed white dwarfs, most of them in the Northern hemisphere. Allowing for a minimum of two current C-enriched DQ identifications to be part of that selection, for a possible total of 4 identifications within the complete 40~pc sample, we estimate a space density of C-enriched DQs of at least $0.75-1.5\times 10^{-5}$~pc$^{-3}$, hence bracketing the C-enriched DQ space density in the range $\rho \approx 0.8-3.0\times10^{-5}$~pc$^{-3}$. At the upper range of space densities, C-enriched DQs could represent nearly half of all DD mergers, or at the lower range no more than a few percent.

\begin{figure}
    \centering
    \includegraphics[viewport=30 275 575 700, clip, width=0.99\columnwidth]{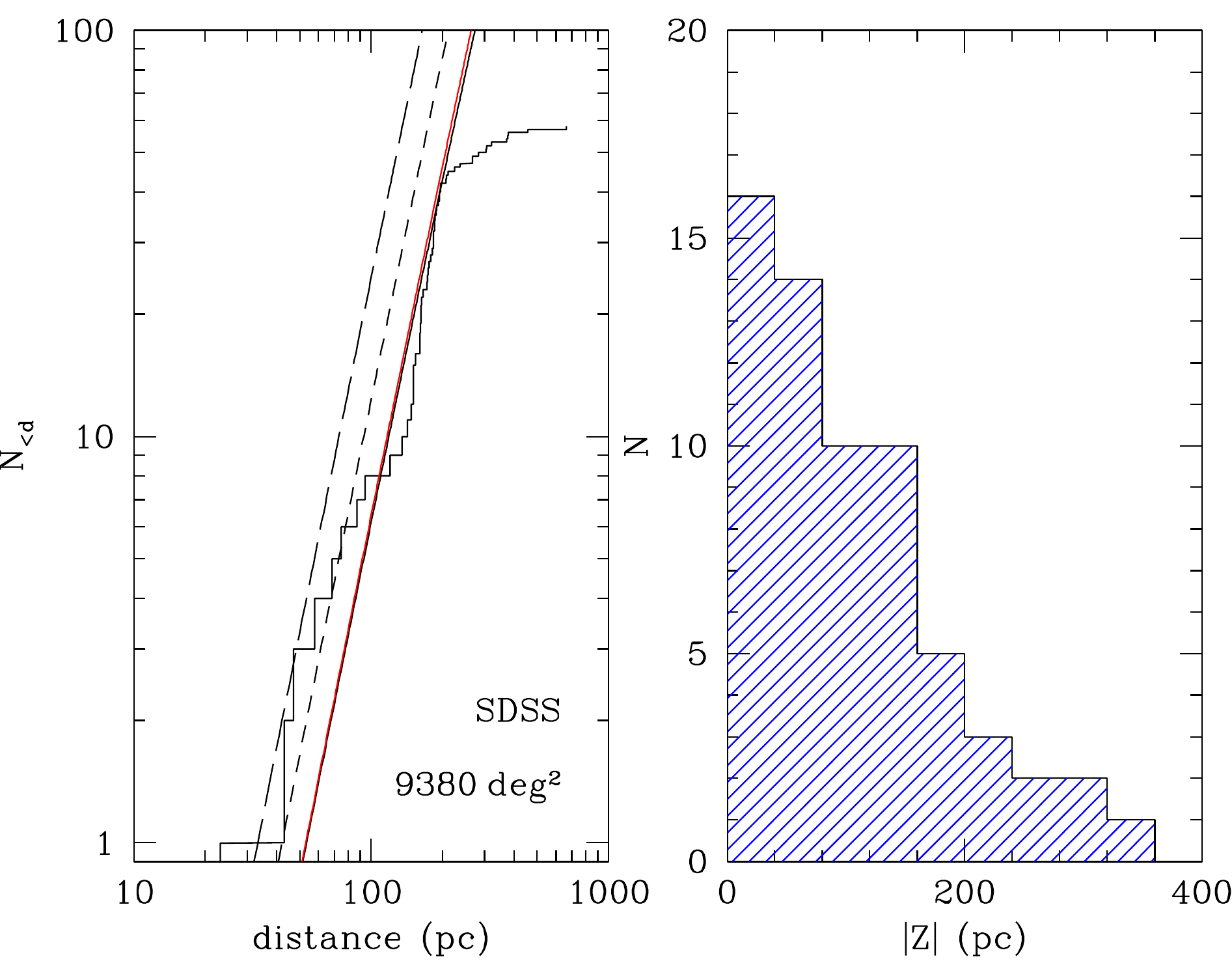}
    \caption{Left panel: Cumulative distance distribution function (histogram) of C-enriched DQs identified in the SDSS with a sky coverage of 9380 degree$^2$. The long-dash, short-dash, and full line curves represent cumulative distributions assuming space density in the Galactic plane of $3\times 10^{-5}$,  $1.5\times 10^{-5}$, and $7.5\times 10^{-6}$~pc$^{-3}$, respectively, assuming a Galactic scale-height of 250~pc and a fractional sky coverage of 0.227 (9380 degree$^2$). The full red line curve represents a cumulative distribution at $\rho_0 = 7.5\times 10^{-6}$~pc$^{-3}$ and $Z_0=350$~pc.
    Right panel: The height,$|Z|$, above the plane of all known massive C-enriched DQs.}
    \label{fig_dist_z}
\end{figure}

Fig.~\ref{fig_dist_z} shows the cumulative distance distribution of C-enriched DQs in the SDSS. Most of these objects were identified in the SDSS which has a current (DR17) sky coverage of 23 percent (9380 degree$^2$), and only five objects out of 63 are not included in the SDSS. The observed SDSS cumulative distribution is compared to calculated distance distributions $N_{<d}$ assuming a space density in the Galactic plane of $\rho_0=3.0$, $1.5$, and $0.75\times 10^{-5}$~pc$^{-3}$, a scale-height $Z_0=250$~pc, and a fractional sky coverage $\Phi=0.227$:
\begin{equation}
    N_{<d}=\Phi\,4\pi\rho_0Z_0^3\Big{[} \frac{1}{2}\Big{(}\frac{d}{Z_0}\Big{)}^2 +\Big{(}\frac{d}{Z_0}+1\Big{)}e^{-d/Z_0}-1 \Big{]}
\end{equation}
The SDSS distribution appears to follow the calculated distribution at $\rho_0=1.5\times10^{-5}$~pc$^{-3}$ up to a distance of 80 pc.
Assuming that all C-enriched DQs were correctly identified over the whole sky and within a distance of 40~pc, i.e., $\rho_0=1.5\times10^{-5}$~pc$^{-3}$, we find that only one fifth of C-enriched DQs have been identified within a distance of 100~pc leaving $\approx40$ bright objects to be discovered, mostly in the Southern hemisphere and outside the SDSS footprint. \citet{Gen2019} showed that
the white dwarf space density decreases with distance using the 100~pc sample of white dwarfs that were selected from the Gaia 2nd Data Release
and that this decrease can be modelled with a scale height of 230~pc. Based on the 2DF sample of distant blue white dwarfs \citet{Vennes2002} showed that the scale height can be as high as 350~pc thereby minimizing the effect of Galactic scale-height of the expected number count of C-enriched DQs. Application of Equation (1) at $Z_0=350$~pc shows that Galactic scale-heights in the 230-350 pc range have a negligible effect on the number count in local surveys ($d\la 100$~pc). We calculated the height above the Galactic plane for all the massive C-enriched DQs and found that they drop off at higher scale heights most likely because of the SDSS survey limit. The majority (87 percent) of stars are found below Z=200~pc (Fig.~\ref{fig_dist_z}).

\section{Origin of C-enriched DQ white dwarfs}\label{origin}

The characteristics of the C-enriched DQs are different from those of ordinary white dwarfs as summarised below. 

\begin{enumerate}

    \item Their optical spectra show notable absorption lines of \ion{C}{i} at 4270.2, 4933.4, 5053.6, 5181.8, 5381.8\AA\ and/or \ion{C}{ii} at 
    4267, 4300, 4370, 4860, 6578, and 6583\AA\, and weak \ion{He}{i} lines. Weak Swan bands are detected at the cooler end of the population ($\approx10\,000$K). The detection of hydrogen in C-enriched DQ white dwarfs is relatively rare, i.e., in less than $\approx20$ percent of the population
    \citep{Dufour2007, DufourHotDQ2008, Coutu2019}, such as in the two massive DQ white dwarfs 
    G35-26 \citep{Thejll1990} and G227-5 \citep{Wegner1985}. An upper limit of $\log{\rm (H/C}) \approx -2.7$ is otherwise achieved among the hottest objects
 from the absence of hydrogen lines in the spectra \citep{DufourHotDQ2008}. Upper limits to the hydrogen abundance in cooler objects have yet to be determined relative to the dominant species, helium or carbon \citep{Coutu2019}.

    \item Their masses are substantially higher. The average mass of H-rich white dwarfs with effective temperatures below 60\,000~K is $0.540$\,M$_\odot$ while that of He-rich white dwarfs below 50\,000\,K is $0.575$\,M$_\odot$ \citep{Bedard2020}. The average mass of the C-enriched DQs is $\langle M \rangle = 1.027\,M_\odot, \sigma_{M} = 0.116\, M_\odot$ (see section \ref{discussion}), thus substantially larger. 
    In fact, their average mass surpasses even that of the magnetic white dwarfs 
    \citep[$\langle M \rangle = 0.87\,M_\odot,\ \sigma_{M} = 0.22\, M_\odot$;][]{Kawka2020}.
    
    This suggests that either the main sequence progenitors of the C-enriched DQs were substantially more massive or that they are the result of mergers.
    
    \item About 70 percent of hot DQs are magnetic \citep{Dufour2013, Dunlap2015} exhibiting field strengths from 0.3~MG up to 18~MG but with the majority of fields between 1 and 4~MG. This is a much higher percentage than among the general white dwarf population \citep[14 - 20 percent in volume-limited samples,][]{Kawka2007}. This is again a strong indicator that stellar merger occurred and was responsible for the generation of their magnetic fields \citep{Tout2008, Garciaberro2012, Wick14, Briggs2015}. 
    
    \item The kinematics characteristics of the C-enriched DQs indicate that they belong to an older population \citep[see Section \ref{kinematics} and][]{Dunlap2015}. Thus these C-enriched DQs are highly unlikely to be the progenies of single massive stars since white dwarfs with masses of $0.8-1.2$\,M$_\odot$ are the descendants of main sequence stars of $3.5-8$\,M$_\odot$ \citep[e.g.,][]{Romero2015} whose evolution to the compact phase is very fast ($\le 0.5$\,Gyrs). Such warm/hot white dwarfs would be confined to the thin Galactic disd. If, instead, they evolved from single stars and became white dwarfs billions of years ago, they would be very cool and dim, unlike the population of the currently observed C-enriched DQs. Therefore, it is much more likely that these DQs are the outcome of merging events many of which occurred with long delay times from the formation of the progenitor binary. 
\end{enumerate}

The points highlighted above strongly support a merger hypothesis for the C-enriched DQ white 
dwarfs \citep{Dunlap2015}. However, not all mergers are expected to produce C-enriched DQ white dwarfs. The 
studies of \citet{Briggs2015} have shown that high field magnetic white dwarfs are the
result of merger events during common envelope evolution with the major contributors coming from low-mass main-sequence stars merging with the degenerate cores of red giant branch (RGB) and asymptotic giant branch (AGB) stars.  The DD mergers only represented a very small fraction of the population of magnetic white dwarfs and populated the high-mass tail of the magnetic white dwarf mass distribution. The latter are the most likely progenitor's candidates of the C-enriched DQs. A second possible channel may consist in the merger of a white dwarf with a naked helium star or in the merger of two naked helium stars. The common property that characterises these two channels is the absence of hydrogen in their envelopes. 

\section{Population synthesis calculations}\label{popsyn}

\begin{figure}
  \centering
\includegraphics[viewport=35 180 460 465,clip,width=0.45\textwidth]{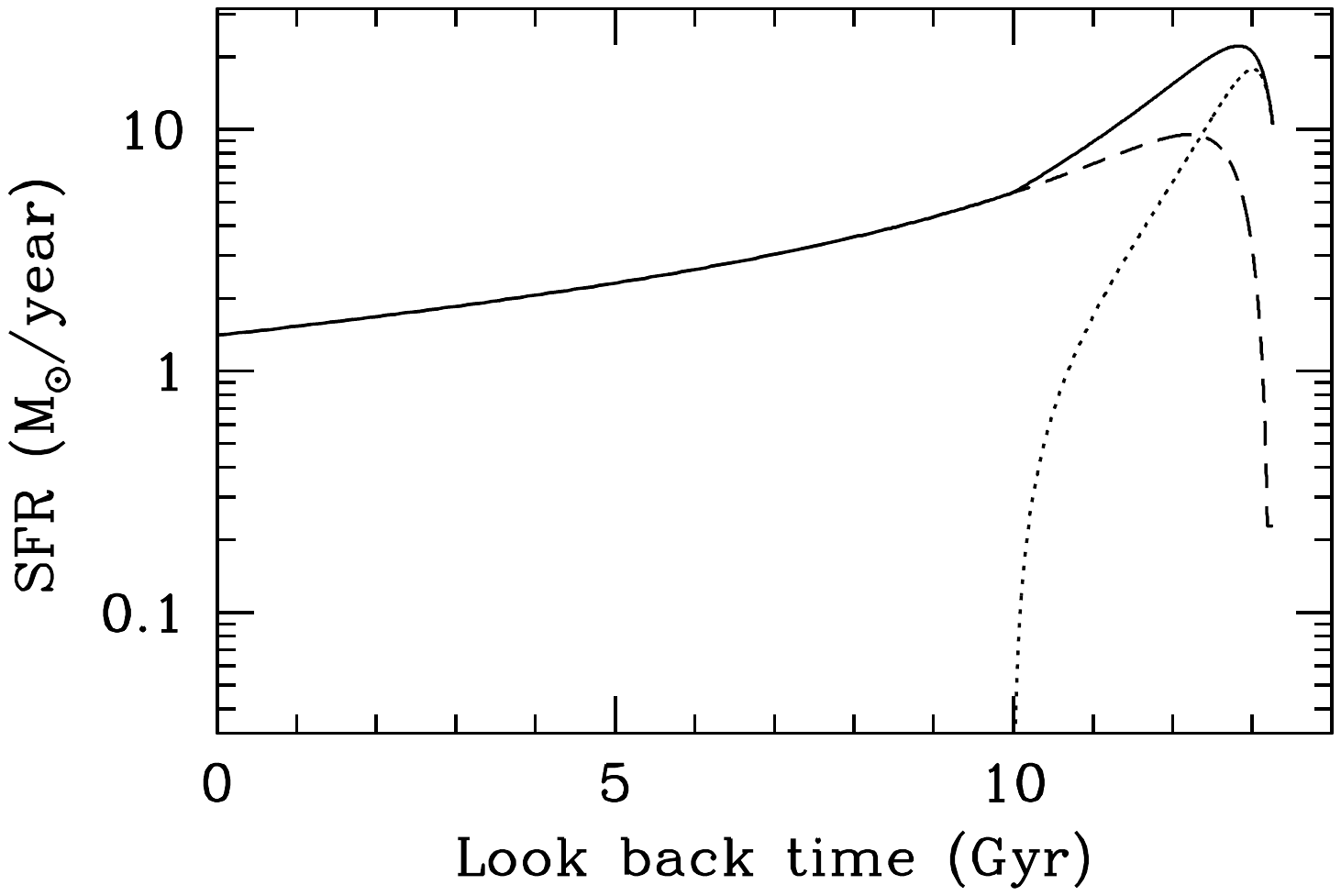}
 \caption{Star formation rate (SFR) of the Galactic disc (dashed line) and bulge (dotted line) with the total SFR shown as a solid curve. The disc's SFR has been used in the present studies. These SFRs yield the observed stellar masses and age distributions of the given components. The SFR reached a maximum about 12.5\,Gyr ago but it is only the Galactic disc that has continued to contribute to star formation up to the present day. }
\label{SFH}
\end{figure}

In order to test the stellar merger hypothesis as an explanation for the origin of the C-enriched DQs, we have used the rapid Binary Stellar Evolution (BSE) code of \citet{Hurley2002} which includes the updates of \citet{Kiel2006}. Common envelope (CE) evolution, as first proposed by \citep{Paczynski1976}, is necessary to explain compact binaries whose size is smaller than the initial radius of the primary star. The outcome of CE evolution is either a merged object or a close binary. Since the processes that govern the CE phase are still not fully understood and the ejection of the envelope may or may not be complete, two parameters are usually introduced. The first is the CE efficiency parameter $\alpha_{\rm CE}$ which was introduced to parametrise the efficiency of the injection of orbital energy into the envelope \citep{Livio1988} and the other is $\lambda_{\rm b}$ which depends on the structure of the donor star and on how tightly bound to the core the envelope is. 
Here we use the BSE option whereby $\lambda_{\rm b}$ is calculated from the detailed stellar evolution models of \citet{Pols1995} obtained with the Cambridge STARS code \citep{Eggleton1971}. An extensive explanation of the nature of this parameter and the range of values it can attain are in \citet{Loveridge2011}.

Since the envelope clearance efficiency is low at small $\alpha_{\rm CE}$'s, the envelope has a longer time to exert a drag on the orbit and consequently the number of stars that merge during CE increases as $\alpha_{\rm CE}$ decreases. Most of these merger events occur during the RGB or AGB phases of the primary star. Those systems that do not coalesce emerge from CE evolution at smaller orbital separations. \citet{Ivanova2013} have shown that the post-CE orbital separation is directly proportional to $\alpha_{\rm CE}$. \citet{Ruiter2011} found that in order to obtain a number of events that is consistent with the predicted rate of SNe\,Ia from the DD merger channel one has to assume complete CE efficiency, that is, $\alpha_{\rm CE}\lambda_{\rm b}=1$. Similarly, 
\citet{Ruiter2019} investigated the various pathways to neutron star formation via the accretion induced collapse (AIC) of oxygen-neon white dwarfs in interacting binaries or via merger induced collapse. They explored their results using two different approaches for CE evolution. In one they have $\alpha_{\rm CE}\lambda_{\rm b}=1$. In the other they keep $\alpha_{\rm CE}=1$ but with the donor binding energy parameter based on the stellar evolution calculations of \citet{Xu2010} and on the evolutionary state of the donor at the onset of CE evolution. They find that the  AIC birthrates are similar in both cases.
In the present work, we also require $\alpha_{\rm CE}=1$, since low $\alpha_{\rm CE}$'s yield too many merged objects with $M\lesssim0.95$. Because of our choice of $\lambda_b$, our simulations are closer to those labelled Model\,2 in \citet{Ruiter2019}.

The masses of the stars, $M_1$ for the primary and $M_2$ for the secondary, are assigned values in the range $0.8-10$\,M$_\odot$ while the initial orbital period at the ZAMS, $P_{\rm i}$, varies in the range $10-10\,000$\,days. The masses of the primary are randomly chosen according to \citet{Kroupa2001} mass function and those of the secondary stars according to a flat mass distribution with $q=M_2/M_1\leq 1$ \citep[e.g.][]{Ferrario2012}. The initial period distribution is assumed to be uniform in the logarithm \citep{Kouwenhoven2007}. The metallicity is near solar taking into consideration that we may slightly underestimate the number of merger events since their rates are higher
at lower metallicity due to lower wind-mass-loss rates \citep{Cote2018}. 

In our population synthesis calculations we have followed the evolution of $10^7$ binaries up to an age $t_{Gal}=13$\,Gyrs. 
From this evolved population we then extracted all single white dwarfs that were the result of either the merger of two white dwarfs or of a white dwarf with a naked helium star or two naked helium stars. All such mergers yield white dwarfs with no hydrogen in their atmospheres. The evolutionary path that leads to these events requires one or two common envelope phases. If the two stars do not coalesce during common envelope evolution, they both evolve to the compact star stage and form a close binary system consisting of two white dwarfs. Because the merger of two white dwarfs is driven by gravitational wave radiation, their merging can be delayed substantially (see section \ref{discussion}). To gain more familiarity with these complex processes, we refer the reader to the thorough review of \citet{Ivanova2013} on binary evolution, on the role that common envelope evolution plays in bringing stars together, and on possible mergers or explosions.

Our population synthesis calculations correspond to a single starburst (one generation of stars). In order to model the currently observed C-enriched DQ population, which is the result of binaries that were born at different Galactic times, we have assigned various birth epochs to the starburst, in agreement with the Galactic disc star formation history (SFH) of \citet{Crocker2017} and shown in Fig\,\ref{SFH}. Briefly, this SFH is given by 
\begin{equation}\label{eSFH}
\log_{10} [SFR + D ] = {\rm max}[Az^2 + Bz + C ,0],
\end{equation}
where $z$ is the cosmological redshift as first proposed by \citet{vanDokkum2013} and \citet{Snaith2014}. This form was then renormalized by \citet{Crocker2017} so that the integrated stellar mass of the disc is $(3.7\pm 0.5)\times 10^{10}$\,M$_\odot$, in agreement with \cite{BlandHawthorn2016}. 
In this context we would like to remark that the star formation history that we have adopted does not take into account a possible inside-out assemblage history of our Galaxy \citep{Xiang2018} and that the effect on the present day merger population remains to be investigated.

This SFH allowed us to scale up the progenitors of the DQs to a number (and thus mass) that makes this subgroup of binaries consistent with the total mass of the Galactic disc. The method consisted in producing many generations of DQs whose progenitor binaries were born at times that were randomly sampled from the SFH of equation (\ref{eSFH}) under the simplifying assumption that there is an equal number of single stars as number of binaries. Each of our merged white dwarf was then assigned a location in the Galaxy in the cylindrical coordinate system ($R$, $\Phi$, $z$) with origin in the Galactic centre. The distribution of stars in the $R$ and $z$ directions were taken to follow exponential laws \citep[e.g.][and references therein]{vanderKruit1982} with radial and vertical scale-lengths of $r_0=2.6\pm0.5$\,kpc and $z_0=800\pm180$\,pc respectively at $R=8$\,kpc \citep[distance of the Sun from the Galactic centre][]{BlandHawthorn2016}. A multiplicative factor $(t/t_{Gal})^{1/2}$, where $t$ is the total age of the star \citep[from the birth of the binary on the main sequence to the present time,][]{Eggleton1989}, has been applied to the distribution in the $z$ direction. This was done to take into consideration the vertical age gradient in the Milky Way disc. Effective temperature and magnitude were assigned to each synthetic object using the tables available at \url{http://www.astro.umontreal.ca/~bergeron/CoolingModels} \citep{Bergeron2011,Bedard2020}.

We have limited our population analysis to objects that have a Gaia G-magnitude, $G$, less than 20. This choice was determined by the study of \citet{Boubert2020} on the completeness of Gaia\,DR2. More specifically, these authors found that over $3 < G < 20$, Gaia is essentially complete and falls from 100 to 0 per cent over $20.0 < G < 21.5$. Thus, all our model data represent a magnitude-limited sample of mergers with $G \le 20$.

The 2nd data release of Gaia revealed an enhancement of massive white dwarfs in a narrow temperature range 
\citep[Q-branch;][]{Gaia2018b} that cannot be
explained with the current cooling models. \citet{che2019} showed that this enhancement could be explained by a delay in the cooling
of massive ($> 1.08$~M$_\odot$) white dwarfs by $^{22}$Ne settling in C/O white dwarfs. This delay could potentially increase the 
cooling age of the cooler white
dwarfs by 8~Gyr. Recently, it was shown that this cooling delay can only occur in white dwarfs with C/O cores and not in 
O/Ne core white dwarfs because crystallization occurs much earlier in the evolution of O/Ne white dwarfs as compared to C/O
core white dwarfs \citep{cam2021,blo2021}. \citet{sch2021} showed the massive white dwarfs that form from the merger of two
C/O white dwarfs end up being O/Ne white dwarfs, and therefore these would not experience this additional delay. The properties of the C-enriched DQs indicate that they may fall into this category.

Before examining the implications of the population syntheses, we first establish the kinematic properties of the population, and in particular its age distribution.

\section{Galactic orbits and stellar populations}\label{kinematics}

\begin{figure}
    \centering
    \includegraphics[viewport=20 160 565 690,clip,width=0.99\columnwidth]{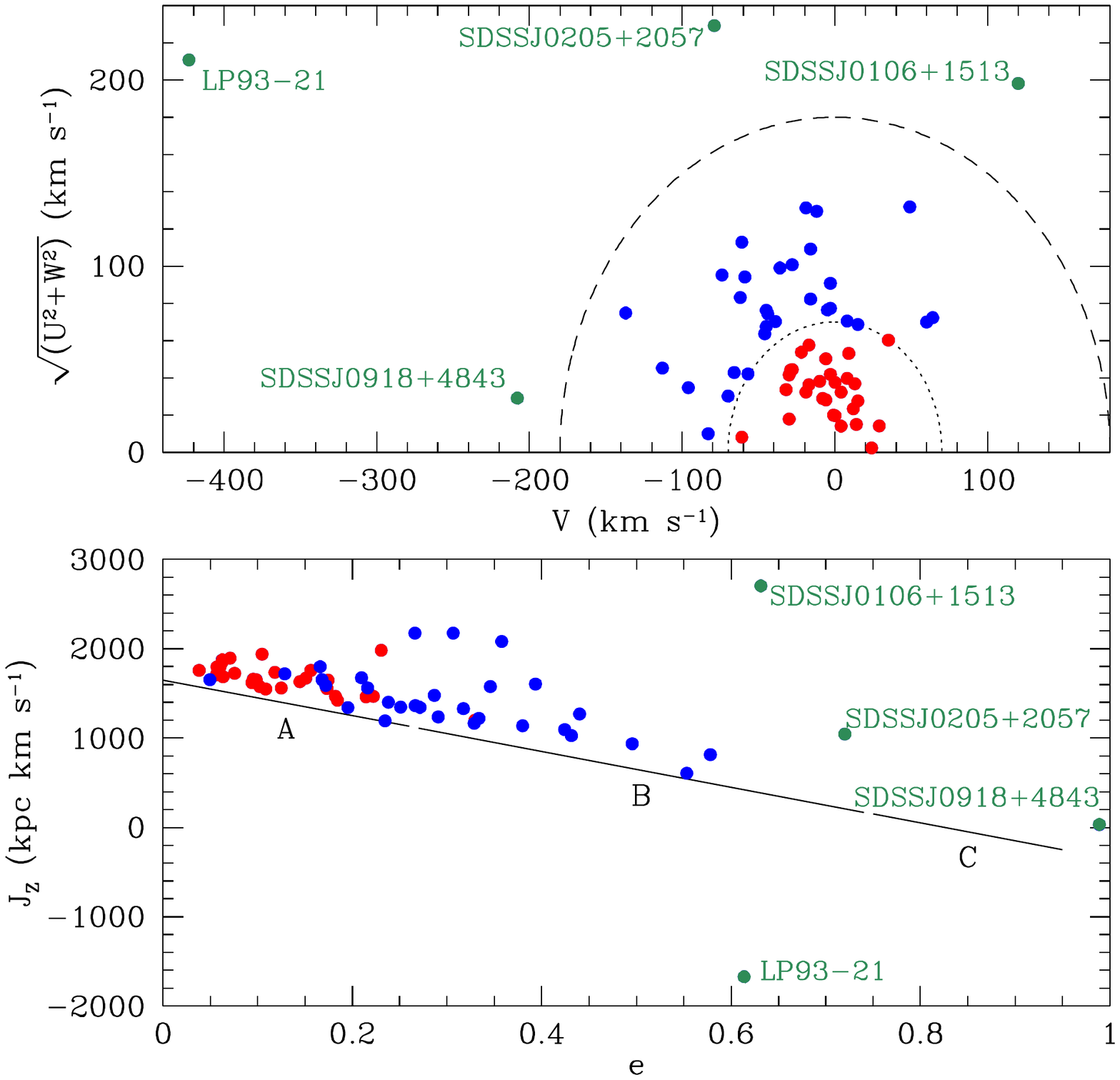}
    \caption{Top panel: $\sqrt{U^2+W^2}$ versus $V$ diagram of the C-enriched DQs. The dotted and short-dashed curves correspond to $v_t = (U^2 + V^2 +W^2)^{1/2}$\,km\,s$^{-1}$ at $70$, and $180$\,km\,s$^{-1}$, respectively, and with respect to the local standard of rest. Bottom panel: Regions $A$, $B$ and $C$ denote thin disc, thick disc, and halo C-enriched DQs.}
    \label{fig_kin}
\end{figure}

We now examine the kinematical properties of the observed sample of C-enriched DQs to establish to which Galactic population they belong. We plot in the top panel of Fig.\,\ref{fig_kin} the Galactic space velocity components as $\sqrt{U^2+W^2}$ versus $V$, where $U$ is positive in the direction of the Galactic centre, $V$ is positive in the direction of the Galactic rotation and $W$ is positive toward the North Galactic pole. The velocities are relative to the local standard of rest.

To a first approximation, stars with a total velocity $v_t= (U^2 + V^2 +W^2)^{1/2}\lesssim 70$\,km\,s$^{-1}$ belong to the thin disc, stars with $70\lesssim v_t \lesssim 180$\,km\,s$^{-1}$ belong to the thick disc \citep{venn2004}, while stars with $v_t\gtrsim 180$\,km\,s$^{-1}$ are likely to be halo objects. There is a likely overlap of thin and thick disc white dwarfs between about $50$ and $70$~km~s$^{-1}$. 

We present the plot of $J_z$ against $e$ in the bottom panel of Fig.\,\ref{fig_kin}. According to \citet{Pauli2006} thin disc stars occupy region $A$ which is characterised by low eccentricities and $J_z$ in the range $1\,600-2\,000$\,kpc\,km\,s$^{-1}$. In region $B$, stars have larger eccentricities and lower $J_z$ and are likely to belong to the thick disc population. Region $C$ is generally populated by halo stars. We can see that the location of the C-enriched DQs in the $J_z$ against eccentricity plot is consistent with that of the $\sqrt{U^2+W^2}$ versus $V$ 
diagram. In particular, we note that there are two DQs that belong to the Galactic halo, 
one of which is LP93-21 \citep{Kawka2020LP93} which is on a retro-grade orbit and the other, SDSS\,J0918+4843, 
appears to have a very
eccentric orbit with near-zero $J_z$. The two other halo candidates identified in the $\sqrt{U^2+W^2}$ versus $V$ diagram 
fall in the thick-disc region in the $J_z$ versus $e$ diagram, however they have a much higher $J_z$ than the other thick-disc candidates. These are likely halo stars since they also have a maximum vertical amplitude $z_{\rm max} > 1.5$~kpc \citep{Martin2017}.

\begin{figure}
    \centering
    \includegraphics[viewport=15 160 565 690,clip,width=0.99\columnwidth]{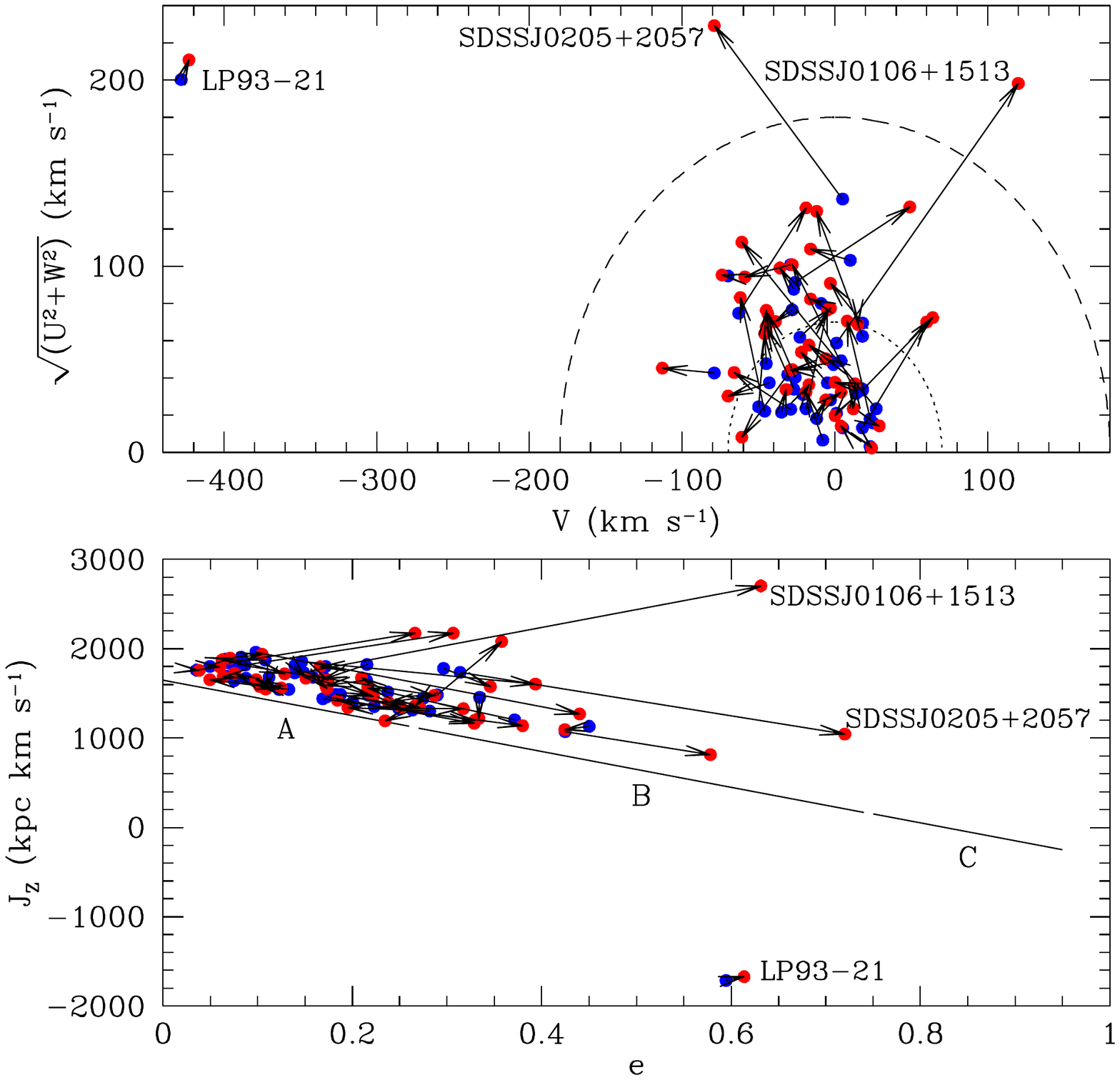}
    \caption{Same as Fig.~\ref{fig_kin}, however here we compare the kinematics of white dwarfs for which we measured a radial 
    velocity (red points) to the same white dwarfs assuming, instead, a zero radial velocity (blue points).}
    \label{fig_vel_comp}
\end{figure}

We have found that about half of the observed 
sample of DQs have kinematic properties that are consistent with those attributed to 
the Galactic thick disc or halo. However this a lower limit because for about one quarter of the sample we had to assume
radial velocities of $0$~km~s$^{-1}$ (see section \ref{kinematics2}). \citet{Pauli2003} suggest that as much as 23 per cent
of thick disc white dwarfs can be misclassified as thin disc when assuming a radial velocity
of $0$~km~s$^{-1}$. We revisit this problem by recalculating the kinematics of the sample for which we have radial velocity measurements and assuming a zero velocity instead.
Fig.~\ref{fig_vel_comp} shows the shifts
in the Galactic velocity components $\sqrt{U^2+W^2}$ versus $V$ and $J_z$ versus $e$ which confirms that the inclusion of the radial 
velocity measurements increases the kinematical age, i.e., it pushes some thin disc stars to the thick disc and thick disc stars
to the halo.

We calculated kinematics for all known C-enriched DQs using Gaia parallaxes and proper motions,
and radial velocities for about two thirds of the sample. Radial velocities for the remaining
third should be acquired to confirm the population kinematics.

\section{Discussion and conclusions}\label{discussion}

\begin{figure}
    \centering
    \includegraphics[viewport=15 150 565 690,clip,width=0.99\columnwidth]{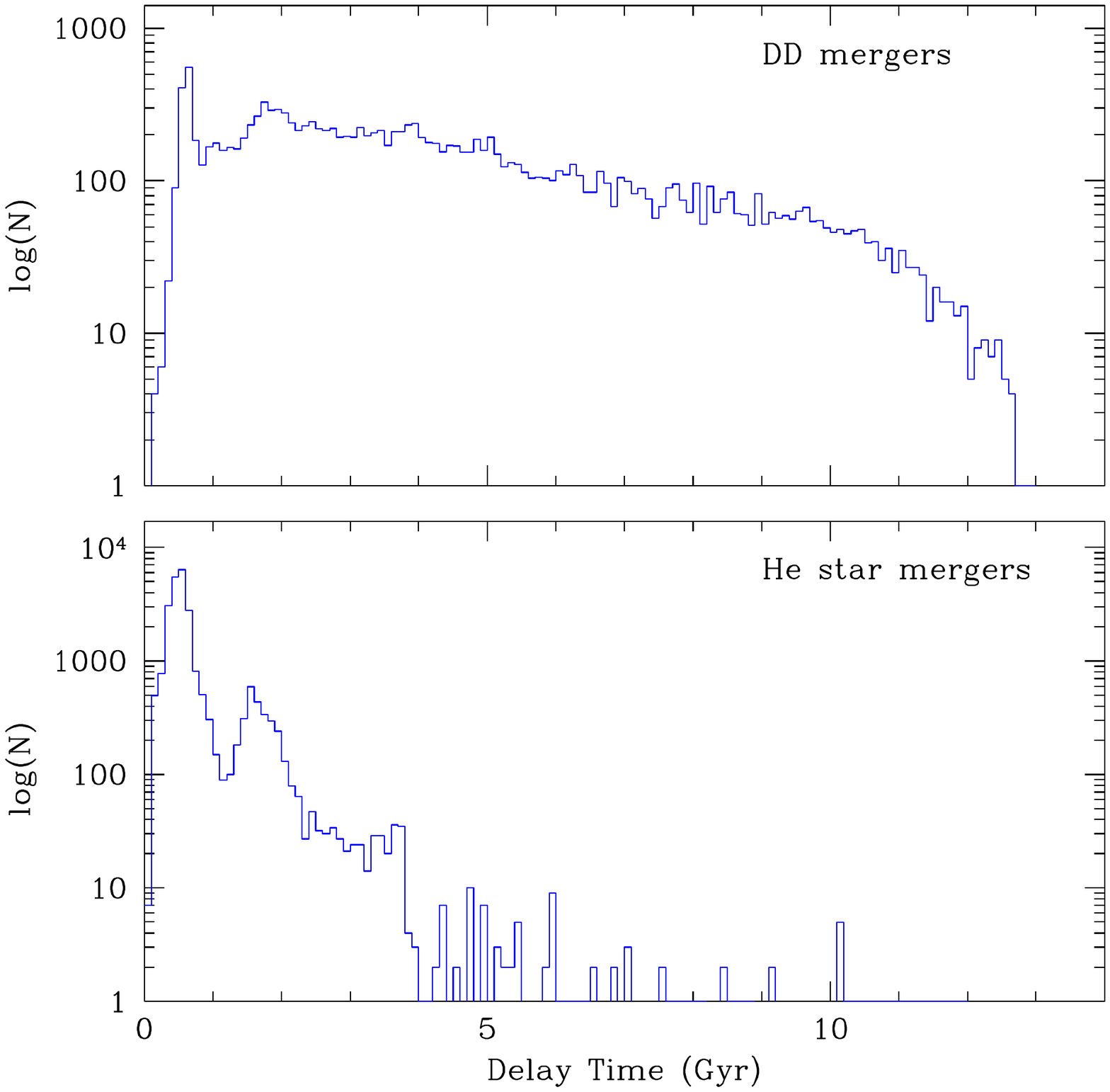}
    \caption{The distribution of delay times of the simulated types of mergers within 1~kpc and for one starburst.}
    \label{fig_delay_times}
\end{figure}

We extracted two samples from the population synthesis calculations. The first contains white dwarfs that formed via DD mergers. The second consists of He star mergers. We show in Fig.\,\ref{fig_delay_times} the type of merger versus delay time, noting that these are simulated data that only pertain to a single starburst. It is obvious that white dwarf-white dwarf progenitors have the longest delay times ranging from a few hundred Myrs to a Hubble time. The delay times of the second sample, instead, are shorter and mostly confined to below a couple of Gyrs. After performing integration over time (see section \ref{popsyn}), we applied the following selection criteria to compare theory to observations. We took all relevant merger products, the DD mergers and He star mergers, within a distance of 200\,pc which is the distance encompassing the majority of known C-enriched DQs (see Fig.~\ref{fig_dist_z}). We also assembled the mergers products brighter than (Gaia magnitude) $G<20$ noting that neither observed sample following such criteria is complete since a number of very dim objects would escape detection even within a distance of 200\,pc. From these selections we extracted the number of objects with a temperature in the range $8\,000 \le T_{\rm eff} \le 25\,000$~K, and, separately, in the range $T_{\rm eff}\ge 25\,000$~K. Table~\ref{table_sim} shows the average and dispersion of the age and mass distributions for the two different samples as well as the number of  objects selected under these criteria. Although the simulated mass distributions appear similar both in their average and dispersion, the projected age distribution of DD mergers corresponds to a much older population than that of He star mergers. This is not surprising since it is entirely consistent with the simulated delay time data portrayed in Fig.\,\ref{fig_delay_times}.

\begin{table*}
\centering
\caption{Statistics of synthetic and observed populations.}
\label{table_sim}
\begin{tabular}{ccccccccc}
\hline
   & & \multicolumn{3}{c}{DD} & & \multicolumn{3}{c}{He-star} \\
\cline{3-5}  \cline{7-9}\\
   & $T_{\rm eff}$ & age/$\sigma_a$  & $M/\sigma_M$ & $N$ & & age/$\sigma_a$  & $M/\sigma_M$ & $N$ \\
   & ($10^3$~K)  & (Gyr) & ($M_\odot$) & & & (Gyr) & ($M_\odot$) &    \\
\hline
$d<200$~pc  & $8-25$      & 6.45/3.43 & 1.131/0.108 & 150 & & 2.36/1.20 & 1.108/0.100 & 291 \\
$d<200$~pc  & $25-\ \ \ $ & 4.55/3.99 & 1.217/0.109 & 10  & & 0.70/0.44 & 1.205/0.132 & 26 \\
$G<20$      & $8-25$      & 5.80/3.38 & 1.098/0.098 & 277 & & 1.84/0.99 & 1.088/0.097 & 512 \\
$G<20$      & $25-\ \ \ $ & 5.75/3.79 & 1.159/0.124 & 126 & & 0.63/0.31 & 1.156/0.108 & 172 \\
            \hline
            \end{tabular}\\
            \begin{tabular}{ccccc}\\
            \hline
               & $T_{\rm eff}$ & age/$\sigma_a$  & $M/\sigma_M$ & $N$\\
                  & ($10^3$~K)  & (Gyr) & ($M_\odot$) & \\
                  \hline
observed   & $8-25$     & 8.51/2.14 & 1.026/0.116 & 63 \\
\hline
 \end{tabular}\\
 \end{table*}
 
\begin{figure}
    \centering
    \includegraphics[viewport=40 165 575 690,clip,width=0.99\columnwidth]{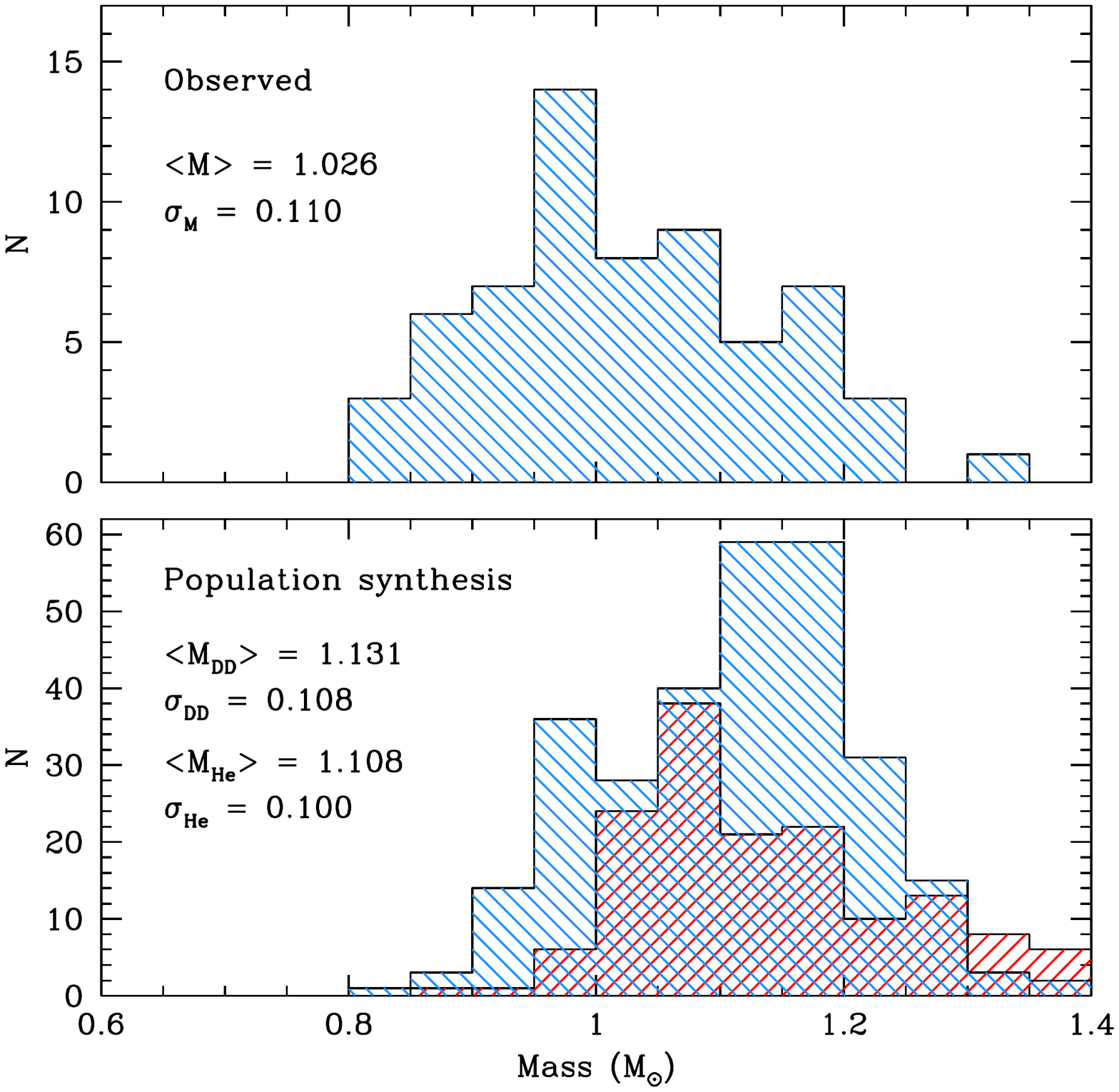}
    \caption{Observed mass distribution of massive and C-enriched DQs (top) compared to the
    mass distribution derived from the $d < 200$~pc population synthesis (bottom). The red and blue histograms in the bottom panel
    show the mass distributions from DD mergers and He star mergers, respectively.}
    \label{mass_distr}
\end{figure}

Fig.~\ref{mass_distr} (top panel) shows the observed mass distribution of the C-enriched DQs. The mean average of the sample is $1.026$~M$_\odot$ with a dispersion of $0.116$~M$_\odot$. This sample is compared to the mass distribution from the DD merger and He star merger population syntheses.
The DD mergers produce more massive white dwarfs, whereas He star mergers can
produce a higher fraction of lower mass white dwarfs when compared to DD
mergers. Both simulated mass distributions peak at slightly higher mass than the observed distribution
with relatively fewer objects than observed at the low end ($0.8\,M_\odot$). The mass distribution of DD mergers simulation peaks at a slightly
lower mass than that of He star mergers. The He star channel appears to form twice as many objects as the DD channel in the $d<200$~pc sample.

\begin{figure}
    \centering
    \includegraphics[viewport=45 160 570 690,clip,width=0.99\columnwidth]{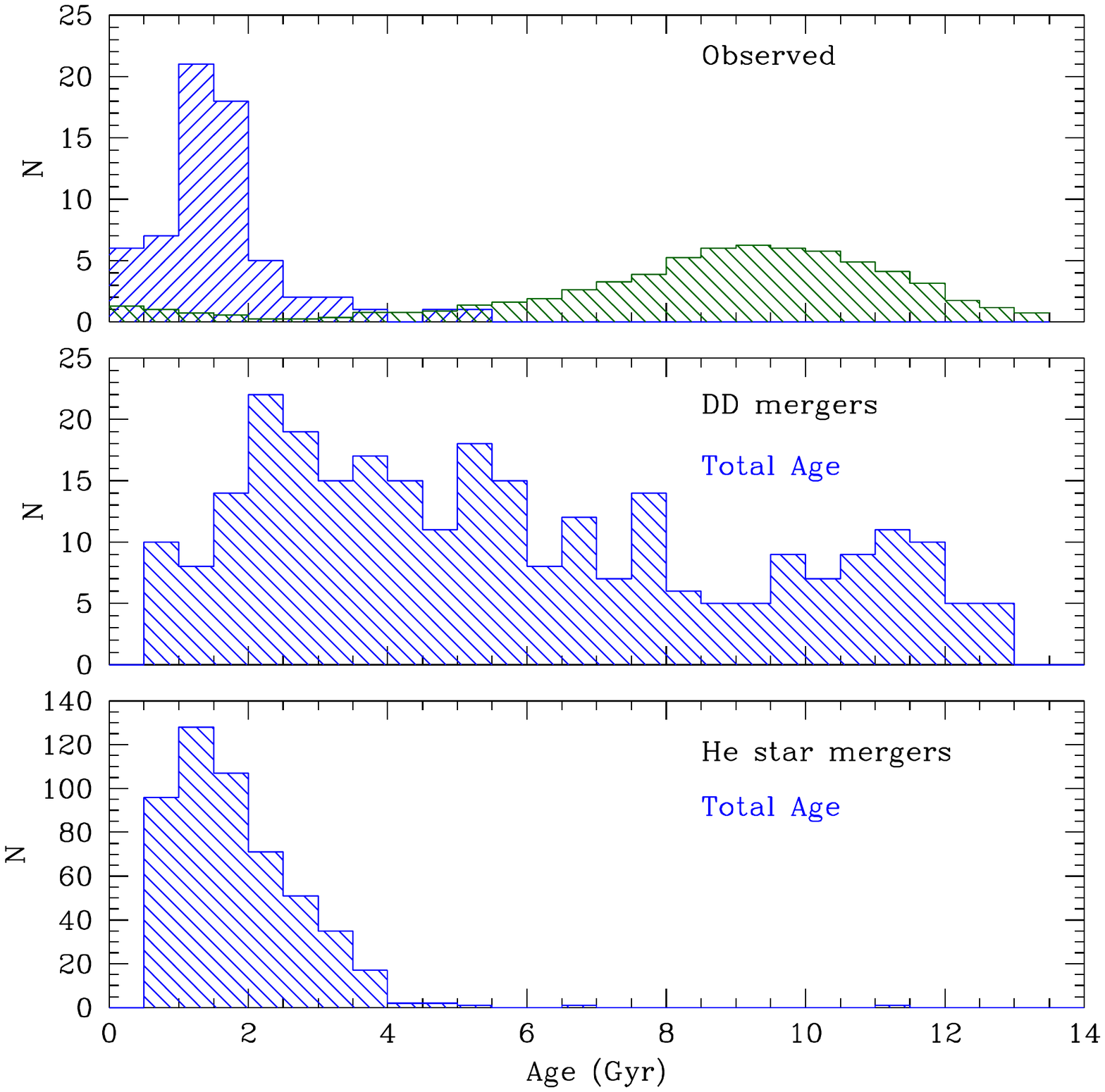}
    \caption{The top panel shows the measured cooling age (+200~Myr) distribution (blue) and eccentricity-determined age distribution (green) of 
    C-enriched DQs. The latter has been smoothed with a 4~Gyr boxcar function to simulate the intrinsic dispersion in the Age-eccentricity relation. The middle panel shows the simulated age distributions for the DD mergers and the
    bottom panel shows the simulated age distributions for the He star mergers. The simulated data apply to the $G < 20$ synthetic sample.}
    \label{fig_age}
\end{figure}

Our population synthesis study does not provide information on space velocity, however we can 
use the observed kinematics, such as the eccentricity of the Galactic orbit, to estimate a 
total age for each white dwarf.
\citet{kor2011} determined the average eccentricity for the thin disc, thick disc and halo, and 
showed that the eccentricity increases with age. We assigned an age of 8~Gyr for the 
thin disc, an age of 10~Gyr for the thick disc \citep{Sharma2019} and an age of 11~Gyr for the halo \citep{Kilic2019}. We fitted
a function (Age $=13.5 - 2.05 e^{-0.5}$, where $e$ is the eccentricity of the Galactic orbit) to these three points that we can apply to the C-enriched DQ white dwarf sample. 
Fig.~\ref{fig_age} (top panel) compares ages of the observed C-enriched DQs assuming single star evolution (blue histogram)
to the age determined from the eccentricity (green histogram). The age assuming single star evolution includes the cooling age of the white dwarf that is based on its mass and temperature added to an average pre-white dwarf lifetime of 200~Myr. In the lower two
panels we show the synthetic populations ($G < 20$) that emerged from DD mergers (middle panel) and He star mergers (bottom panel). The age distribution of objects emerging from the DD channel is in qualitative agreement with the kinematic age of the C-enriched DQs, although our simulations suggest that there should be a larger number of C-enriched DQs than currently observed. As mentioned earlier, the C-enriched white dwarfs would only be a subset of all merger products alongside other merger candidates such as ultra-massive magnetic white dwarfs.  It demonstrates that the much longer lifespan of C-enriched DQs relative to their apparent (cooling) age is the result of binary evolution and interaction in the form of DD mergers.
The age of the He star channel products is much lower and corresponds to a thin Galactic disc population rather than the observed thick disc or halo populations. 

Results obtained from the $G<20$ sample in the population synthesis show a large excess of He star merger products relative to the DD merger products. Again the age distribution of these objects does not match the observed distribution. 

Neither observed samples, $d<200$~pc and $G<20$, are complete but the simulations show that many more objects should be identifiable at current and future survey limits and that He star mergers should dominate in the thin Galactic disc. However, only DD merger products take the appearance of the kinematically old population of C-enriched DQs. 

Because our population synthesis study does not provide information on space velocity we have not addressed the excess of C-enriched DQs with a transverse velocity $>70$\,km\,s$^{-1}$ in the Q-branch region of the H-R diagram that was attributed by \citet{che2019} to some additional cooling delay mechanisms. However, we have concluded that the products of mergers are likely to produce 
O/Ne core white dwarfs which do not experience additional cooling since crystallization occurs at higher temperatures than those of white dwarfs on the Q-branch. Therefore, if white dwarfs produced from mergers do experience additional cooling as they pass through the Q-branch, it cannot be through $^{22}$Ne settling and another mechanism is required.

We know that one of the evolutionary channels
leading to Type Ia SNe, used as standard candles to measure cosmological distances, consists in the merger of two white dwarfs. We can therefore state that these massive C-enriched DQs are failed Type Ia SNe, as first noted by \citet{Dunlap2015}. We have also shown that
these mergers are rare events and that only a few C-enriched DQs are observed with an 
estimated space density that is between 0.2 to 0.7 percent of the local space
density of white dwarfs. Nonetheless, they may constitute from a few to about $50$ percent of all DD merger products.

Our study shows that this population of white dwarfs is old, with nearly half of the observed objects having kinematic properties consistent with those of stars belonging to the Galactic thick disc and halo. Our population synthesis results largely support these findings and are compatible with a population of white dwarfs descending from  the merger of two white dwarfs. We found that the merger of stars whose envelope was stripped of hydrogen during common envelope evolution (He star mergers) would leave remnants much younger than actually observed. Note that the simulated distributions were not actually fitted to observed distributions; the predicted and observed population numbers may differ by more than a factor of two but are generally of the same order of magnitude. The population synthesis predicts a large number of very hot white dwarfs
($25\,000 < T_{\rm eff} \lesssim 110\,000$) that resulted from DD mergers.
They would represent 15 percent of DD mergers in a volume-limited survey ($d<200$~pc), and up to 75 percent in a magnitude-limited survey ($G<20$). These white dwarfs will most likely have carbon rich atmospheres not unlike the hottest known objects in the C-enriched population. Few such objects are known, but the ultra-hot, massive DZQ H1504+65 and cooler siblings \citep{Wer2015} which show a mixed carbon-oxygen atmosphere are emerging as possible candidates. The observed trend in carbon abundance with temperature in these likely merger products remains to be explained.

To conclude, we note that binary white dwarfs are sources of low-frequency gravitational waves. Therefore, some of the progenitors of these merging binaries will be detectable with the space-based gravitational wave observatory LISA, which is an European Space Agency-led mission, scheduled to launch in the early 2030's. Whilst most binary white dwarfs are invisible in the electromagnetic spectrum, LISA will be able to `hear' thousands of them millions of years before they merge.

\section*{Acknowledgements}

LF and SV would like to express their gratitude for the hospitality of the staff at the International Centre for Radio Astronomy Research. We thank D.T. Wickramasinghe for useful discussions. This study is partly based on observations made with ESO telescope at the
La Silla Paranal Observatory under programmes 097.D-0694 and 097.D-0063 and 090.D-0536. We thank M.S. Bessell for sharing with us the spectrum of J2140$-$3637.
Funding for the SDSS IV has been provided by the Alfred P. Sloan Foundation, the U.S. Department of Energy Office of Science, and the Participating Institutions. SDSS-IV acknowledges support and resources from the Center for High-Performance Computing at
the University of Utah. The SDSS web site is www.sdss.org. This work presents results from the European Space Agency (ESA) space mission Gaia. Gaia data are being processed by the Gaia Data Processing and Analysis Consortium (DPAC). Funding for the DPAC is provided by national institutions, in particular the institutions participating in the Gaia Multi-Lateral Agreement (MLA). The Gaia mission website is https://www.cosmos.esa.int/gaia. The Gaia archive website is https://archives.esac.esa.int/gaia.

\section*{Data Availability}

The SDSS spectra are available publicly from the SDSS Archive (https://www.sdss.org/). The FORS2 and UVES spectra are from the author (AK).




\bibliographystyle{mnras}
\bibliography{DQ} 

\clearpage

\appendix

\section[]{Two new carbon-enriched DQ white dwarfs}

\begin{figure}
\centering
\includegraphics[viewport=15 155 570 700,clip,width=0.99\columnwidth]{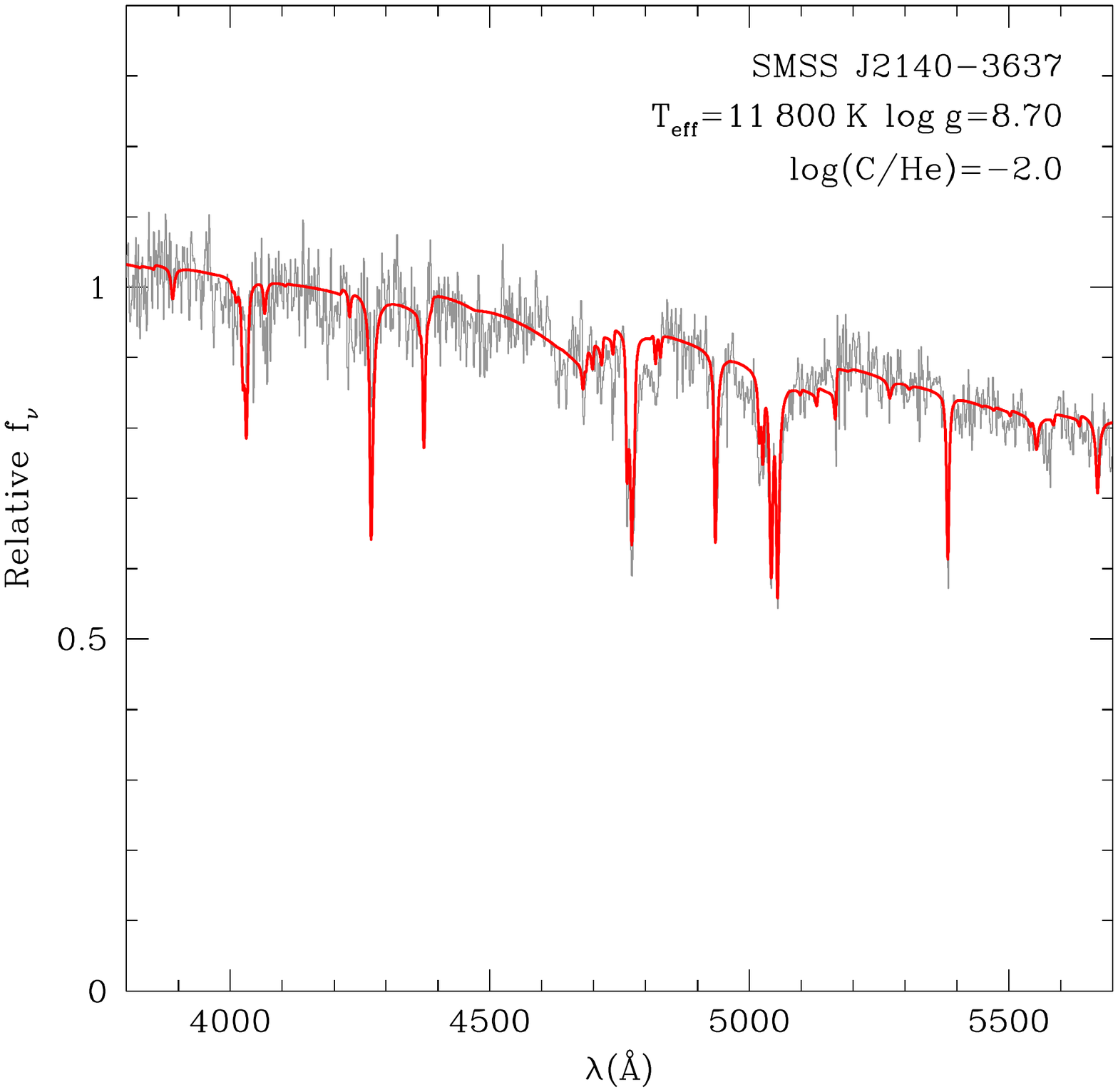}
\caption{Optical spectra (in grey) and best-fit spectral syntheses (in red) of the newly identified C-enriched DQs WD~J2140$-$3637.}\label{fig_2140}
\end{figure}

\begin{figure}
\centering
\includegraphics[viewport=15 155 570 700,clip,width=0.99\columnwidth]{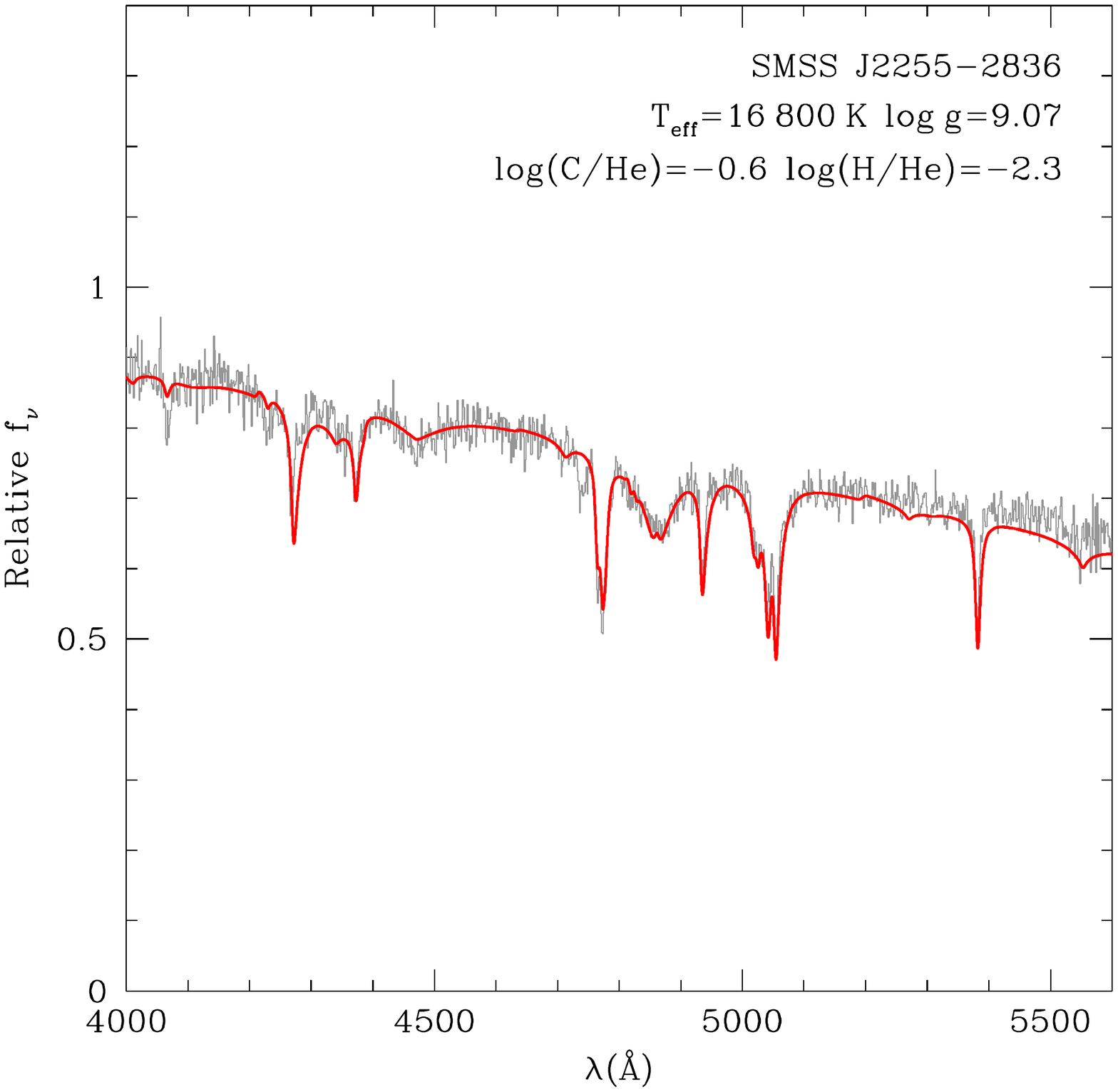}
\caption{Optical spectra (in grey) and best-fit spectral syntheses (in red) of the newly identified C-enriched DQs WD~J2255$-$2836.}
\label{fig_2255}
\end{figure}

We included in this work two newly identified carbon-enriched white dwarfs. The objects were identified in the SkyMapper survey of white dwarfs (Vennes et al., in preparation). The two objects, WD~J2140$-$3637 and WD~J2255$-$2836, lie along the abundance versus temperature relation (Fig.~\ref{fig_abun}) at $T_{\rm eff}=11\,800\pm1\,000$~K and $\log{\rm C/He}\approx-2$, and $T_{\rm eff}=16\,800\pm1\,800$~K and $\log{\rm C/He}\approx-0.6$, respectively. The cooler object, WD~J2140$-$3637, shows C$_2$ Swan bands and atomic carbon lines. Also, along with approximately one in five C-enriched DQs in the \citet{Coutu2019} sample, WD~J2255$-$2836 shows contamination by hydrogen ($\log{\rm H/He}\approx-2.3$). These objects are among new DQ identifications in the Southern hemisphere complementing the SDSS Northern hemisphere coverage. Fig.~\ref{fig_2140} shows the SSO2.3m/WiFeS optical spectrum of WD~J2140$-$3637 and Fig.~\ref{fig_2255} shows the VLT/FORS spectrum  of WD~J2255$-$2836, along with their best-fit spectral syntheses. We employed mixed H/He/C convective model atmospheres. A complete description of these two objects and models will be presented elsewhere (Vennes et al., in preparation).

\section[]{Atmospheric and kinematic parameters}

Table~\ref{tab_prop_massive} lists all known DQ white dwarfs with a surface composition enriched in carbon relative to normal DQ white dwarfs; its members have higher than average mass and effective temperatures extending up to 25\,000~K. We list the distance calculated from the Gaia parallax and the stellar parameters (effective temperature, surface gravity, mass, and carbon abundance) obtained from the literature or determined in this work (see Section 2).  

Table~\ref{tbl_kin} lists the radial velocities measured in this work along with the Galactic velocity components $U, V, W$, the Galactic orbital eccentricity $e$, and the angular momentum along the Z-axis $J_Z$ (see Sections 2.1 and 5).

\begin{table*}
\centering
\caption{Atmospheric parameters of C-enriched DQ white dwarfs.}
\label{tab_prop_massive}
\begin{tabular}{llcccccc}
\hline
    & Name & Distance  & $T_{\rm eff}$ & $\log{g}$ & $M$  & $\log{{\rm C/He}}$ & Reference \\
     &      &    (pc)       &     (K)           &  (c.g.s.) &   ($M_\odot$)   &         &    \\
\hline
J0005$-$1002$^a$ & SDSSJ000555.90$-$100213.3, PHL~657 & 150.7 &  19018 &   8.80 &  1.10 &  \ 2.00 & 1 \\
J0019$+$1847 & SDSSJ001908.63$+$184706.0  & 150.3 &  10280 &   8.55 &  0.93 &  $-$2.84 & 2 \\
J0045$-$2336 & G268-40                    &  47.2 &  10500 &   8.65 &  1.00 &  $-$2.70 & 3,4 \\
J0106$+$1513$^a$ & SDSSJ010647.92$+$151327.8  & 372.0 &  23430 &   8.50 &  0.93 &  \ 1.00 & 4,5 \\
J0205$+$2057 & G35-26                     &  85.4 &  16150 &   9.04 &  1.20 &  \ 3.00 & 6 \\
J0236$+$2503 & SDSSJ023633.74$+$250348.9  & 177.7 &  13376 &   8.70 &  1.03 &  $-$1.58 & 7 \\
J0236$-$0734$^a$ & SDSSJ023637.42$-$073429.5  & 663.0 &  24400 &   9.07 &  1.22 &  \ 2.00 & 4 \\
J0243$+$0101 & SDSSJ024332.77$+$010111.1, WD0240$+$008               & 182.4 &   8225 &   8.63 &  0.99 &  $-$4.23 & 7 \\
J0807$+$1949 & SDSSJ080708.48$+$194950.7  & 171.3 &  13501 &   8.78 &  1.08 &  $-$1.24 & 7 \\
J0818$+$0102 & SDSSJ081839.23$+$010227.5  & 266.8 &  24483 &   8.33 &  0.81 & \ 2.00 & 2 \\
J0852$+$2316 & SDSSJ085235.43$+$231644.3  & 185.9 &  11099 &   8.61 &  0.97 &  $-$3.18 & 7 \\
J0856$+$4513 & SDSSJ085626.94$+$451336.9  & 205.9 &   9484 &   8.51 &  0.91 &  $-$3.27 & 7 \\
J0859$+$3257 & SDSSJ085914.63$+$325712.1, G47-18                     &  23.1 &   9486 &   8.45 &  0.87 &  $-$3.52 & 7 \\
J0901$+$5751 & SDSSJ090157.93$+$575135.9, WD0858+580                 & 153.8 &  13576 &   8.76 &  1.07 &  $-$1.99 & 7 \\
J0918$+$4843 & SDSSJ091830.27$+$484323.0  & 184.4 &   9203 &   8.80 &  1.09 &  $-$3.72 & 7 \\
J0919$+$0236 & SDSSJ091922.22$+$023604.5, WD0916$+$028               & 159.9 &  11319 &   8.61 &  0.98 &  $-$2.85 & 7 \\
J0936$+$0607 & SDSSJ093638.07$+$060710.0  & 161.7 &  11013 &   8.61 &  0.97 &  $-$3.07 & 7 \\
J0958$+$5853 & SDSSJ095837.00$+$585303.0  & 175.1 &  15444 &   8.95 &  1.16 &  $-$0.50 & 2 \\
J1036$+$6522$^a$ & SDSSJ103655.38$+$652252.0, WD1033+656             & 175.3 &  15500 &   8.83 &  1.12 &  $-$1.00 & 8,4 \\
J1040$+$0635 & SDSSJ104052.40$+$063519.7  & 283.7 &  13882 &   8.40 &  0.84 &  $-$2.08 & 7 \\
J1045$+$5904 & SDSSJ104559.14$+$590448.2, LP93-21                    &  57.7 &   9730 &   8.90 &  1.14 &  $-$2.73 & 9 \\
J1049$+$1659 & SDSSJ104906.61$+$165923.6  & 194.1 &  12799 &   8.92 &  1.15 &  $-$1.64 & 7 \\
J1058$+$2846 & SDSSJ105817.66$+$284609.3  & 162.6 &   9422 &   8.49 &  0.89 &  $-$3.60 & 7 \\
J1100$+$1758 & SDSSJ110058.03$+$175806.9  & 150.9 &  12367 &   8.76 &  1.07 &  $-$1.28 & 7 \\
J1104$+$2035$^a$ & SDSSJ110406.68$+$203528.7  & 173.4 &  23476 &   8.60 &  0.99 &  \ 2.00 & 1 \\
J1113$+$0146 & SDSSJ111341.33$+$014641.7  &  43.1 &   5961 &   8.71 &  1.03 &  $-$5.14 & 10 \\
J1133$+$6331 & SDSSJ113359.94$+$633113.3, WD1131$+$637               & 194.1 &  11517 &   8.57 &  0.95 &  $-$2.68 & 7 \\
J1140$+$0735 & SDSSJ114059.85$+$073530.1  & 160.8 &  10651 &   8.56 &  0.94 &  $-$3.36 & 7 \\
J1140$+$1824 & SDSSJ114006.35$+$182402.3  &  94.2 &   9656 &   8.35 &  0.81 &  $-$3.54 & 7 \\
J1148$-$0126 & SDSSJ114851.68$-$012612.7, WD1146$-$011               &  68.2 &   9680 &   8.46 &  0.88 &  $-$3.48 & 7 \\
J1153$+$0056 & SDSSJ115305.54$+$005646.2  & 165.3 &  21650 &   9.40 &  1.39 &  \ 2.00 & 4,5 \\
J1200$+$2252 & SDSSJ120027.73$+$225212.9  & 378.6 &  21880 &   8.50 &  0.92 &  \ 2.00 & 2 \\
J1203$+$6451 & SDSSJ120331.89$+$645101.4, WD1200$+$651               &  87.2 &  12359 &   8.77 &  1.07 &  $-$1.59 & 7 \\
J1209$+$5355 & SDSSJ120936.50$+$535525.7  & 236.0 &  11721 &   8.53 &  0.92 &  $-$2.71 & 7 \\
J1215$+$4700 & SDSSJ121510.64$+$470011.0  & 160.6 &  13230 &   8.87 &  1.13 &  $-$2.00 & 7 \\
J1225$+$0959 & LP495-79                   &  84.4 &  11100 &   8.60 &  0.97 &  $-$2.60 & 11,4 \\
J1328$+$5908$^a$ & SDSSJ132858.19$+$590851.0, WD1327$+$594           & 147.7 &  18755 &   9.01 &  1.19 &  \ 3.00 & 6 \\
J1331$+$3727 & SDSSJ133151.38$+$372754.8  & 134.9 &  16741 &   9.03 &  1.16 &  \ 0.41 & 2 \\
J1332$+$2355 & SDSSJ133221.56$+$235502.1  & 210.6 &  14205 &   8.70 &  1.03 &  $-$1.76 & 7 \\
J1337$-$0026$^a$ & SDSSJ133710.19$-$002643.7  & 304.5 &  22711 &   8.66 &  1.02 &  \ 1.50 & 1 \\
J1339$+$5036 & SDSSJ133940.53$+$503612.8  & 182.4 &  11680 &   8.62 &  0.98 &  $-$2.20 & 2 \\
J1341$+$0346 & SDSSJ134124.28$+$034628.7  & 196.1 &  13765 &   8.76 &  1.07 &  $-$2.18 & 7 \\
J1400$-$0154 & SDSSJ140051.57$-$015414.4, 2QZJ140051.6$-$015413      & 142.3 &   9394 &   8.69 &  1.02 &  $-$3.56 & 7 \\
J1402$+$3818$^a$ & SDSSJ140222.26$+$381848.9  & 320.5 &  17232 &   8.61 &  0.99 &  $-$0.50 & 1 \\
J1426$+$5752$^a$ & SDSSJ142625.70$+$575218.4  & 306.3 &  18809 &   8.72 &  1.04 &  \ 2.00 & 2 \\
J1428$+$3238 & SDSSJ142812.54$+$323817.7  & 161.8 &  10718 &   8.56 &  0.94 &  $-$3.20 & 7 \\
J1434$+$2258 & SDSSJ143437.82$+$225859.5  & 191.9 &  14575 &   8.75 &  1.06 &  $-$1.14 & 7 \\
J1435$+$5318 & SDSSJ143534.01$+$531815.0  & 196.0 &  15167 &   8.85 &  1.12 &  $-$1.91 & 7 \\
J1444$+$0434 & SDSSJ144407.25$+$043446.7, WD1441$+$047               & 180.5 &   9813 &   8.44 &  0.87 &  $-$3.47 & 7 \\
J1448$+$0519 & SDSSJ144854.80$+$051903.5  & 120.3 &  15966 &   8.94 &  1.16 &  \ 0.30 & 2 \\
J1452$+$6020 & SDSSJ145236.57$+$602036.3, WD1451$+$605               & 224.9 &  12572 &   8.65 &  1.00 &  $-$1.73 & 7 \\
J1455$+$4209 & SDSSJ145524.89$+$420910.8  & 266.4 &  14288 &   8.78 &  1.08 &  $-$1.48 & 7 \\
J1542$+$4329 & SDSSJ154248.67$+$432902.4  & 184.4 &   9799 &   8.49 &  0.89 &  $-$3.61 & 7 \\
J1555$+$3219 & SDSSJ155539.51$+$321914.1  & 189.1 &   9195 &   8.59 &  0.96 &  $-$3.74 & 7 \\
J1615$+$4543 & SDSSJ161531.71$+$454322.4  & 455.5 &  20940 &   8.62 &  1.00 &  \ 1.74 & 10,4 \\
J1622$+$1849 & SDSSJ162205.12$+$184956.7  & 187.7 &  16693 &   9.13 &  1.16 &  $-$0.08 & 2 \\
J1622$+$3004 & SDSSJ162236.13$+$300454.5  &  74.7 &  16131 &   8.93 &  1.15 &  $-$0.10 & 2 \\
J1728$+$5558 & SDSSJ172856.19$+$555823.0, G227-5                     &  47.1 &  14453 &   8.90 &  1.14 &  $-$1.37 & 7 \\
 \hline
 \end{tabular}\\
 \end{table*}
 
 \begin{table*}
\centering
\contcaption{Atmospheric parameters of massive DQ white dwarfs.}
\label{tab_prop_massive_cont}
\begin{tabular}{llcccccc}
\hline
    & Name & Distance & $T_{\rm eff}$ & $\log{g}$ & $M$ & $\log{{\rm C/He}}$ & Reference \\
     &      &    (pc)       &     (K)  &  (c.g.s.) &   ($M_\odot$)   &         &    \\
\hline
J2140$-$3637 & SMSSJ214023.58$-$363757.5   & 39.8 & 11800 &  8.70  & 1.02  & $-$2.00 & 4 \\
J2200$-$0741$^a$ & SDSSJ220029.09$-$074121.5 & 207.2 & 21271 &   8.64 &  1.01 &  \ 2.00 & 1 \\
J2250$+$1240 & SDSSJ225000.22$+$124019.8  & 182.5 &  9801 &   8.50 &  0.90 &  $-$3.66 & 7 \\
J2255$-$2836 & SMSSJ225523.30$-$283649.6  & 153.7 & 16800 &  9.07  & 1.22  & $-$0.60 & 4 \\
J2348$-$0942 & SDSSJ234843.30$-$094245.2  & 375.6 & 21550 &   8.54 &  0.93 &  \ 2.00 & 4,5 \\
  \hline
 \end{tabular}\\
 References: (1)~\citet{Hardy2018} (2)~\citet{Koester2019} (3)~\citet{Koester1982} (4)~This work 
 (5)~\citet{Dufour2008}  (6)~\citet{Leggett2018} (7)~\citet{Coutu2019} (8)~\citet{williamsetal13-1} 
 (9)~\citet{Kawka2020LP93} (10)~\citet{Blouin2019} (11)~\citet{demartino2007}\\
 $^a$ Confirmed to be magnetic.
\end{table*}

\begin{table*}
\centering
\caption{Kinematics of C-enriched DQs.}
\label{tbl_kin}
\begin{tabular}{ccccccc}
\hline
           & $v_r$        & $U$           & $V$           &  $W$          & $e$ & $J_z$ \\
           & (km~s$^{-1}$) & (km~s$^{-1}$) & (km~s$^{-1}$) & (km~s$^{-1}$) & & (kpc km~s$^{-1}$) \\
\hline
J0005$-$1002 &   $-10$ &  $-49$ &    $-6$ &    $11$ &   $0.174$ &  $1648$ \\
J0019$+$1847 &   $-38$ &  $109$ &   $-16$ &    $-6$ &   $0.346$ &  $1575$ \\
J0045$-$2336 &    $76$ &  $-25$ &   $-45$ &   $-72$ &   $0.194$ &  $1341$ \\
J0106$+$1513 &   $228$ &  $-45$ &   $120$ &  $-193$ &   $0.631$ &  $2703$ \\
J0205$+$2057 &  $-195$ &  $227$ &   $-79$ &    $31$ &   $0.720$ &  $1044$ \\
J0236$+$2503 &    $-1$ &   $14$ &     $4$ &     $0$ &   $0.038$ &  $1757$ \\
J0236$-$0734 &    $28$ &   $20$ &    $13$ &   $-31$ &   $0.063$ &  $1876$ \\
J0243$+$0101 &   ($0$) &   $29$ &   $-96$ &   $-19$ &   $0.496$ &   $938$ \\
J0807$+$1949 &    $93$ &  $-41$ &  $-113$ &    $19$ &   $0.578$ &   $814$ \\
J0818$+$0102 &   $137$ & $-109$ &   $-61$ &    $29$ &   $0.440$ &  $1269$ \\
J0852$+$2316 &    $53$ &   $-9$ &   $-22$ &    $53$ &   $0.109$ &  $1549$ \\
J0856$+$4513 &   ($0$) &  $-15$ &    $15$ &   $-23$ &   $0.062$ &  $1852$ \\
J0859$+$3257 &     $6$ &  $-17$ &    $12$ &   $-16$ &   $0.060$ &  $1796$ \\
J0901$+$5751 &  $-113$ &  $126$ &   $-12$ &   $-30$ &   $0.393$ &  $1606$ \\
J0918$+$4843 &   ($0$) &  $-29$ &  $-208$ &     $1$ &   $0.989$ &    $33$ \\
J0919$+$0236 &    $47$ &   $68$ &   $-59$ &    $65$ &   $0.334$ &  $1221$ \\
J0936$+$0607 &    $-7$ &    $5$ &   $-17$ &   $-36$ &   $0.102$ &  $1576$ \\
J0958$+$5853 &   $-24$ &  $-49$ &   $-16$ &   $-66$ &   $0.172$ &  $1590$ \\
J1036$+$6522 &  $-103$ &   $83$ &   $-28$ &   $-57$ &   $0.287$ &  $1479$ \\
J1040$+$0635 &   ($0$) &  $-16$ &    $-1$ &   $-12$ &   $0.057$ &  $1709$ \\
J1045$+$5904 &    $15$ & $-203$ &  $-423$ &    $57$ &   $0.614$ & $-1672$ \\
J1049$+$1659 &   $-12$ &   $14$ &    $29$ &     $2$ &   $0.105$ &  $1940$ \\
J1058$+$2846 &    $26$ &   $81$ &   $-74$ &    $50$ &   $0.424$ &  $1097$ \\
J1100$+$1758 &   $-18$ &   $23$ &    $-6$ &   $-16$ &   $0.098$ &  $1652$ \\
J1104$+$2035 &    $36$ &  $-16$ &   $-19$ &    $28$ &   $0.125$ &  $1559$ \\
J1113$+$0146 &   ($0$) &   $16$ &   $-57$ &   $-39$ &   $0.291$ &  $1235$ \\
J1133$+$6331 &   ($0$) &  $-11$ &  $-137$ &    $74$ &   $0.553$ &   $606$ \\
J1140$+$0735 &    $37$ &   $44$ &   $-17$ &    $37$ &   $0.173$ &  $1554$ \\
J1140$+$1824 &    $67$ &  $-43$ &   $-45$ &    $52$ &   $0.251$ &  $1346$ \\
J1148$-$0126 &    $69$ &   $15$ &   $-66$ &    $40$ &   $0.329$ &  $1167$ \\
J1153$+$0056 &   $-90$ &  $-70$ &   $-19$ &  $-111$ &   $0.216$ &  $1559$ \\
J1200$+$2252 &   ($0$) &  $-38$ &   $-10$ &    $-3$ &   $0.145$ &  $1630$ \\
J1203$+$6451 &   $-58$ &   $29$ &   $-70$ &    $-8$ &   $0.380$ &  $1136$ \\
J1209$+$5355 &   ($0$) &  $-11$ &   $-30$ &    $14$ &   $0.182$ &  $1468$ \\
J1215$+$4700 &   $-79$ &   $55$ &   $-44$ &   $-50$ &   $0.271$ &  $1344$ \\
J1225$+$0959 &    $5$  &  $-44$ &   $-29$ &     $3$ &   $0.222$ &  $1466$ \\
J1328$+$5908 &   ($0$) &  $-37$ &   $-30$ &    $19$ &   $0.215$ &  $1464$ \\
J1331$+$3727 &    $61$ &   $19$ &    $-5$ &    $74$ &   $0.050$ &  $1655$ \\
J1332$+$2355 &   $-17$ &   $17$ &     $0$ &   $-10$ &   $0.064$ &  $1685$ \\
J1337$-$0026 &    $57$ &   $40$ &     $8$ &    $58$ &   $0.129$ &  $1721$ \\
J1339$+$5036 &   ($0$) &  $-59$ &    $35$ &   $-12$ &   $0.231$ &  $1981$ \\
J1341$+$0346 &    $45$ &  $-61$ &   $-36$ &    $78$ &   $0.238$ &  $1402$ \\
J1400$-$0154 &   ($0$) &   $-8$ &   $-83$ &    $-6$ &   $0.431$ &  $1026$ \\
J1402$+$3818 &   ($0$) &  $-12$ &    $14$ &     $9$ &   $0.057$ &  $1798$ \\
J1426$+$5752 &     $2$ &    $2$ &    $24$ &     $1$ &   $0.071$ &  $1893$ \\
J1428$+$3238 &   $-70$ &   $54$ &    $-3$ &   $-73$ &   $0.168$ &  $1655$ \\
J1434$+$2258 &    $13$ &   $17$ &   $-32$ &    $29$ &   $0.184$ &  $1423$ \\
J1435$+$5318 &    $78$ &  $-16$ &    $60$ &    $68$ &   $0.266$ &  $2175$ \\
J1444$+$0434 &   ($0$) &   $38$ &     $8$ &   $-11$ &   $0.118$ &  $1734$ \\
J1448$+$0519 &   $-57$ &  $-63$ &   $-46$ &    $-9$ &   $0.317$ &  $1328$ \\
J1452$+$6020 &    $32$ &  $-59$ &    $-3$ &    $50$ &   $0.210$ &  $1673$ \\
J1455$+$4209 &   ($0$) &   $24$ &    $-8$ &    $16$ &   $0.094$ &  $1620$ \\
J1542$+$4329 &   ($0$) &   $53$ &     $9$ &    $-1$ &   $0.156$ &  $1759$ \\
J1555$+$3219 &   ($0$) &  $-23$ &    $-3$ &    $35$ &   $0.095$ &  $1658$ \\
J1615$+$4543 &   $-27$ &  $-37$ &     $0$ &    $-6$ &   $0.152$ &  $1672$ \\
J1622$+$1849 &   $-23$ &   $59$ &   $-39$ &   $-38$ &   $0.266$ &  $1364$ \\
J1622$+$3004 &   $-55$ &   $-8$ &   $-61$ &     $0$ &   $0.329$ &  $1201$ \\
J1728$+$5558 &   $-38$ &  $-44$ &   $-28$ &     $6$ &   $0.219$ &  $1478$ \\
J2140$-$3637 &    $-3$ &   $51$ &    $15$ &    $46$ &   $0.166$ &  $1799$ \\
J2200$-$0741 &    $82$ &   $62$ &    $64$ &   $-37$ &   $0.307$ &  $2177$ \\
J2250$+$1240 &    $99$ &  $-68$ &    $49$ &  $-113$ &   $0.358$ &  $2081$ \\
J2255$-$2826 &   $-90$ &  $-14$ &   $-62$ &    $82$ &   $0.234$ &  $1193$ \\
J2348$-$0942 &    $18$ &   $29$ &     $4$ &   $-14$ &   $0.076$ &  $1725$ \\
\hline
\end{tabular}
\end{table*}

\bsp	
\label{lastpage}
\end{document}